# Bridging Superconductors with UN Development Goals: Perspectives and Applications

Edimar A. S. Duran[1], Alfonso Pulgar[1], Rodolfo Izquierdo[1], Diana M. Koblischka[2], Anjela Koblischka-Veneva[2,3], Michael R. Koblischka[2,3], and Rafael Zadorosny[1,*]

[1]Universidade Estadual Paulista (UNESP), Faculdade de Engenharia, Caixa Postal 31, 15385-000, Ilha Solteira, SP, Brazil
[2]SupraSaar, Auf der Ochsenweide 31, 66133 Saarbrücken, Germany
[3]Experimental Physics, Saarland University, 66041 Saarbrücken, Germany
[*]Author to whom correspondence should be addressed: rafael.zadorosny@unesp.br

September 25, 2024

## Abstract

Superconductors exhibit remarkable properties such as zero resistivity and diamagnetism at the boiling temperature of liquid hydrogen (20 K) and even above the boiling temperature of liquid nitrogen (77 K), making them promising candidates for various applications including electrical en- gines, energy generation, storage, and high-tech devices like single photon detectors. In this overview, we explore the correlation between ceramic superconductors and the United Nations (UN) Develop- ment Goals, emphasizing their potential impact on sustainable development. Through bibliometric analysis, we underscore the significance of ceramic superconductors in addressing global challenges outlined by the UN. Additionally, we discuss the application of supermagnets and second-generation tapes in healthcare systems, particularly in magnetic resonance imaging (MRI) devices for diagnostic imaging. Electric superconducting motors offer a clean alternative to highly polluting diesel engines in maritime transportation and superconducting wires/cables enable effective transport of energy on large scale as well as in industrial structures. Magnetic levitation technology holds promise for de- veloping zero-emission public transportation systems, and magnetic separation with strong magnets will contribute to solve the microplastic pollution. The combination of superconductivity with the planned hydrogen economy further offers new possibilities to bring superconductivity to common applications. At the nanoscale, superconducting nanowire single photon detectors (SNSPDs) enable real-time monitoring of environmental health, exemplified by applications in plant physiology, and superconducting qubits provide the best-engineered structures for quantum computers. However it is still crordwork to works with superconductors UN develepoment goals.

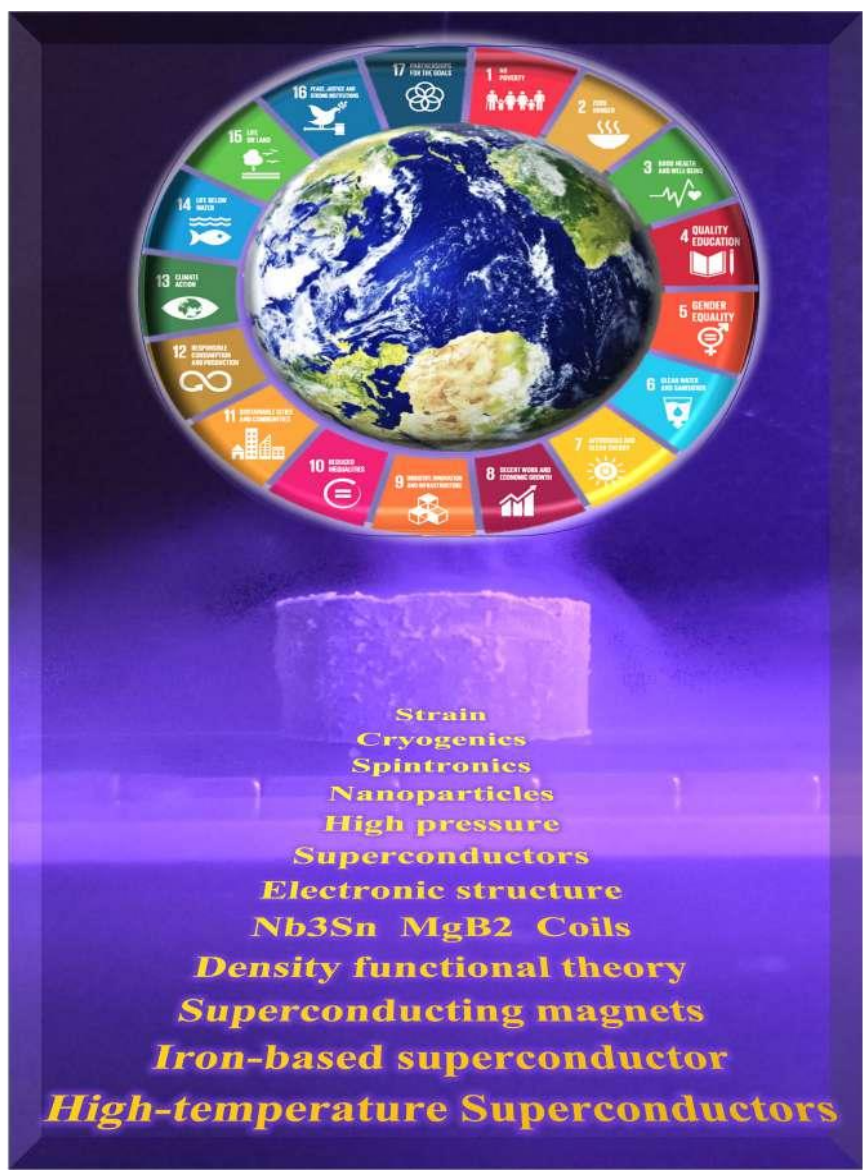

## Highlights

- A comprehensive bibliometric analysis reveals the global evolution of superconductivity research from 2015 to 2023.
- The role of superconductors in advancing UN Sustainable Development Goals is systematically explored.
- Integration of superconductors with hydrogen technologies presents innovative pathways for sustainable energy solutions.
- Addressing interdisciplinary challenges to harness superconductors' potential in achieving UN SDGs.

# List of Abbreviations

| Abbreviation | Description |
| --- | --- |
| AC | Alternating current |
| BSCCO | Bismuth strontium calcium copper oxides Cumulative |
| CAN | number of articles |
| CAS | Chinese Academy of Sciences |
| COP | Conference of the Parties Critical |
| $H_C$ | magnetic field Current density |
| $J_C$ | Direct current |
| DC | United States Department of Energy Electron |
| DOE | paramagnetic resonance Energy Storage |
| EPR | Technologies |
| EST FC | Field cooling |
| FCL FCS | Fault current limiter |
| FESS | Fluorescence Correlation Spectroscopy Flywheel |
| FLASH | energy storage systems |
| GCC GDP | Fast Low Angle Shot Global |
| GHG | climate change Gross |
| GHGs | domestic product |
| GLT GW | Greenhouse gas Greenhouse |
| HTSs | gases Ginzburg-Landau |
| IPCC | Theory Global warming |
| $I_c$ | High-temperature superconductors |
| LTSs | Intergovernmental Panel on Climate Change Limiting |
| Maglev | critical current |
| MDGs | Low-temperature superconductors |
| MEG | Magnetic levitation |
| $MgB_2$ | Millennium Development Goals |
| MRI | Magnetoencephalography |
| $Nb_3Sn$ | Magnesium Diboride |
| NbN NbTi | Magnetic Resonance Imaging |
| NHMFL | Niobium-tin |
| NMR | Niobium nitride |
| PARTER | Niobium-titanium |
| PLD | National High Magnetic Field Laboratory Nuclear |
| QD | Magnetic Resonance |
| QS | Advanced Radiotherapy Research Platform Pulsed |
| REBCO | laser deposition |
| SC | Quantum Device Quantum |
| SCH | sensors |
| SDGs | Rare Earth Barium Copper Oxide Superconductors |
| SMES | Series-connected resistive/superconducting hybrid Sustainable |
| SQUID | Development Goals |
| $T_c$ | Superconducting Magnetic Energy Storage |
| UNCED | Superconducting Quantum Interference Device |
| UNFCCC | Transition temperatures |
| VOC | United Nations Conference on Environment and Development United Nations |
| WHO | Framework Convention on Climate Change Volatile organic compounds |
| WoS | World Health Organization Clarivate |
| XRD | Analytics Web of Science X-ray |
| YBCO | diffractometry |
| IBS | Yttrium barium copper oxides Iron-based superconductors |

# 1 Introduction

Today, humanity has achieved unprecedented population and economic growth. According to the Bank World's data, world population and gross domestic product (GDP) surpassed 8 bn and US$100 $^{th}$ in 2022 [1]. Simultaneously, human development, economic growth, and industrial production demand energy consumption at an unsustainable rate. For example, the global total energy demand was 8795 Mtoe in 1990 and 14080 Mtoe in 2017 [2]. There has been a significant shift in energy production in the last decades [2]. In addition, despite the efforts in clean energy, so far, fossil fuel-based energies dominate the global energy market (e.g., 83% of global consumption was from fossil origin in 2020) [3, 4].

The most alarming consequence of the excessive consumption of fossil fuels is the global climate change (GCC) promoted by human activity. This led to a change towards a new geological stage, i.e., the Anthropocene era [5], which is characterized by humans' actions [6], changing the planet equilibrium and tracing a persistent trajectory toward global warming (GW) [7], unprecedented increase in anthropogenic levels of primary pollutants released from fossil fuel combustion such as carbon dioxide ($CO_2$) [8], volatile organic compounds (VOC) [9], sulfur oxides ($SO_x = SO_2 + SO_3$) [10], and nitrogen oxides ($NO_x = NO + NO_2$) [11].

In particular, $CO_2$ is a greenhouse gas (GHG) that plays a vital role in maintaining the Earth's temperature, and is primarily responsible for GW. The level and trajectory of the greenhouse gases (GHGs) are seen to be increasing in a quasilinear relationship (see Fig 2 in ref. [12]) when plotted as the level of $CO_2$ against global temperature rise [13, 14]. This trend is truly alarming; in the last six years, about 35 Gtoe $CO_2$ were produced per year, mainly due to carbonaceous fuel combustion and human activities [15], and about 43 Gtoe $CO_2$ is projected to be produced in 2040 [16]. Based on these levels of $CO_2$ emitted at a pace exceeding the worst projection of the Intergovernmental Panel on Climate Change (IPCC) [17], the mean surface temperature in 2100 is projected to be 4 °C higher than the 1986–2005 average [18]. Hence, our climate situation threatens the future stability of the Earth ecological system as a whole, and thus, the survival of human race.

On the other hand, human activities during the Anthropocene have induced profound alterations in the biosphere, atmosphere, and oceans [19–21]. Climate change, driven by human actions, increasingly has widespread adverse impacts on nature and various populations. Vulnerable populations who have historically contributed the least to current climate change are disproportionately affected, and the so- cioeconomic differences between developed and developing countries have reached unsustainable levels [22–24]. For example, approximately 3.3 to 3.6 bn people live in contexts that are highly vulnerable to climate change [25].

Climate change has been regarded as the single largest global health challenge in the $21^{st}t$ century by affecting the physical environment and ecosystem and their interactions with human beings. The health outcomes range from premature deaths caused by natural disasters to communicable diseases due to deteriorated hygiene and over-proliferation of pathogens [18]. These health impacts of climate change have been well studied and extensively documented [26–29]. Thus, the interconnected nature of climate change's effects across ecological, environmental, sociopolitical, and socioeconomic domains makes it a terrible and global problem [30].

The first time climate change was addressed at the highest political level occurred at the United Nations Conference on Environment and Development (UNCED), also known as the 'Earth Summit', held in Rio de Janeiro, Brazil, from 3-14 June 1992 [31]. Next, the Millennium Development Goals of the UN (MDGs), all by the target date of 2015, marked a historic and effective method of global mobi- lization to achieve a set of critical social priorities worldwide. They expressed widespread public concern about poverty, hunger, disease, unmet schooling, gender inequality, and environmental degradation [32]. Unfortunately, the report entitled "A Life of Dignity for All" presented at the UN, on 25 September 2013, concluded that despite the efforts made towards achieving the MDGs, the results were insufficient, and it was agreed to hold a high-level Summit in September 2015 to adopt a new set of Goals building on the achievements of the MDGs (i.e., the Sustainable Development Goals) [33, 34]. In particular, the Sustainable Development Goals (SDGs) were released and ratified an agreement by the UN to improve global sustainability by 2030 [35]. The SDGs are made up of 17 goals, and 169 targets, and include 300 indicators covering all aspects of sustainability and are an ambitious step toward actionable targets for sustainable development covering all aspects of sustainability and all sectors of society (see Table 1) [36].

Table 1 shows the highlights of each SDG, and according to Fonseca et al. [37], mapping the re- lationships is a challenging topic. However, the correlations published by the Fonseca's group confirm that SDG 01 (poverty elimination) and SDG 03 (good health and well-being) have synergetic relation- ships with the remaining fifteen goals. SDG 07 (affordable and clean energy) has significant relationships with almost all other goals (e.g., SDG 01 (no poverty), SDG 02 (zero hunger), SDG 03 (good health and well-being), SDG 08 (decent work and economic growth), SDG 13 (climate action), and a moderate negative correlation with SDG 12 (responsible consumption and production), which emphasizes the need to improve energy efficiency, increase the share of clean and renewable energies and improve sustainable consumption patterns worldwide. Thus, SDG 02 is strongly associated with environmental regulations and policies that are nourished by SDG 11 (sustainable cities and communities), SDG 12 (responsible consumption and production), SDG 13 (climate action), SDG 14 (life below water), SDG 15 (life on land), and SDG 17 (partnerships for the goals). In this review, we will use some of these relationships between SDGs as a guide in our discussions. Other interesting approaches to relationships between objectives can be reviewed in Refs. [38–40].

Table 1: Top twenty most active countries based on number of articles per Corresponding Author's Countries

| Sustainable Development Goals (SDGs) | Highlights according to Fonseca *et al.* [37] | SDG Icons |
|---|---|---|
| SDG 01. No poverty | End poverty in all its forms, everywhere. | 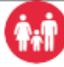 |
| SDG 02. Zero hunger | End hunger, achieve food security and improved nutrition, and promote sustainable agriculture. | 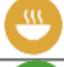 |
| SDG 03. Good health and well-being | Ensure healthy lives and promote well-being for all at all ages. | 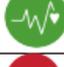 |
| SDG 04. Quality education | Ensure inclusive and equitable quality education and promote lifelong learning opportunities for all. | 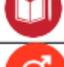 |
| SDG 05. Gender equality | Achieve gender equality and empower all women and girls. | 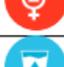 |
| SDG 06. Clean water and sanitation | Ensure available and sustainable management of water and sanitation for all. | 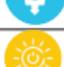 |
| SDG 07. Affordable and clean energy | Ensure access to affordable, reliable, sustainable and modern energy for all. | 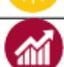 |
| SDG 08. Decent work and economic growth | Promote sustained, inclusive and sustainable economic growth, full and productive employment, and decent work for all. | 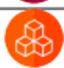 |
| SDG 09. Industry, innovation, and infrastructure | Build resilient infrastructure, promote inclusive and sustainable industrialization, and foster innovation. | 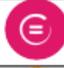 |
| SDG 10. Reduced inequalities | Reduce inequality within and among countries. | 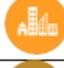 |
| SDG 11. Sustainable cities and communities | Make cities and human settlements inclusive, safe, resilient and sustainable | 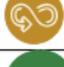 |
| SDG 12. Responsible consumption and production | Ensure sustainable consumption and production patterns | 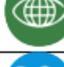 |
| SDG 13. Climate action | Take urgent action to combat climate change and its impacts. | 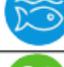 |
| SDG 14. Life below water | Conserve and sustainably use the oceans, seas and marine resources for sustainable development. | 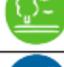 |
| SDG 15. Life on land | Protect, restore and promote sustainable use of terrestrial ecosystems, sustainably manage forests, combat desertification, and halt and reverse land degradation, and halt biodiversity loss. | 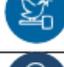 |
| SDG 16. Peace, justice and strong institutions | Promote peaceful and inclusive societies for sustainable development, provide access to justice for all, and build effective, accountable and inclusive institutions at all levels. | 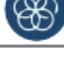 |
| SDG 17. Partnerships for the goals | Strengthen the means of implementation and revitalize the global partnership for sustainable development. | 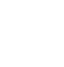 |

The SDGs provide a unique opportunity to create a unified framework for furthering human prosperity in an era of growing evidence of rising global environmental risks. Science can provide independent guidance on goal and target formulations to help increase the likelihood of meeting policymakers' stated sustainability [41]. In this way, the Paris Agreement (PA) constitutes a perfect gear between two generally diametrically opposed sectors, political and scientific. PA was signed by 195 nations at the 21$^{st}$ Conference of Parties (COP21) of the United Nations Framework Convention on Climate Change (UNFCCC) in December 2015. It intended to commit the parties to keep global temperatures from rising more than 2 °C by 2100 with an ideal target of keeping temperature rise below 1.5 °C [42, 43]. Despite some negative indicators of implementing the PA [44–46], it is currently widely accepted that The Climate Policy Based on the Paris Agreement marked unprecedented progress in the combat against climate change [47–49].

Considering the abovementioned issues, the imperative for compliance with the SDGs and PA becomes evident. This fact is crucial to avert a climate catastrophe and the well-being of society, the economy, and global health. To fulfill these complex tasks requires the development of new materials at an unprecedented pace. Bontempi et al. [50] using a thorough bibliometric analysis classified the sustainable materials according to their contributions to achieving the SDGs in the following categories: chemicals (33%), composites (11%), novel materials for pollutants sequestration (8%), bio-based and food-based materials (10%), materials for green building (8%), and materials for energy (29%). Chu et al. [51] classified the sustainable materials according to the following categories: (i) materials for gas separation and storage, (ii) materials for thermal energy conversion and manipulation, (iii) materials for power electronics, (iv) microbial energy conversion. Zhu and collaborators [52] classified the sustainable materials according to their ability as ultra-strength capacity in nanoparticles, nanowires, nanotubes, nanopillars, thin films, and nanocrystals. The general approach in these reviews is based on describing the materials according to synthesis or bulk manufacturing, chemical composition, and physicochemical properties. Still, those reviews need to include the details of their applications. Here, we only briefly summarize this fact, but a more in-depth analysis can be find elsewhere [53–58].

In particular, superconducting materials provide bold solutions to various sustainable development challenges [59]. These materials exhibit the superconducting phenomenon in a region below the interdependent values of critical temperature $T_c$, upper critical magnetic field $H_{c2}$, and critical current density $J_c$. The maximum $T_c$ for low-temperature superconductors (LTSs) is in the range of 23 K to 40 K, whereas for high-temperature superconductors (HTSs) is greater than the boiling temperature of nitrogen (77 K) [60, 61]. There are few reviews that address sustainability and superconductors issues, Gielen et al. [62] highlighted some examples of cases in which new materials play a fundamental role in the search for technically and economically viable solutions for sustainable technologies. Among the classification described by them, the superconductors for low-cost and long-range energy transmission were highlighted. Muralidhar et al. [63] recently published some progress in the mass production of bulk $(Gd_{0.33}Y_{0.13}Er_{0.53})Ba_2Cu_3Oy$ "(Gd,Y,Er)123" and $MgB_2$ systems, and its impact to stop climate change and reach sustainable development. Nishijima et al. [59] reviewed the role of superconductivity in global environmental challenges; the authors emphasized some areas where superconductivity could have an impact on the environment, e.g., water purification, power distribution and storage, low-environmental impact transport, environmental sensing, and so on.

Keeping in mind the balance between human and technological development, while responsibly using natural resources and considering our interactions with all living beings on Earth, we present this review on the $8^{th}$ anniversary of the UN Sustainable Development Goals. This review aims to understand how the academic community is exploring research alternatives to propose actions or articles related to this theme, particularly regarding superconducting applications. In addition, we can highlight that as a guide for review readers after the introduction section, Section 3 provides a brief bibliometric analysis focused on keywords such as superconductors, applications, and SDGs. Section 7 presents a bibliometric study that shows the indicators during the SDGs years. In Section 5, the dynamics of keywords and topics during the SDGs era are described and discussed. Section 8 presents some superconducting applications, considering the results obtained from the bibliometric analysis. Finally, in the conclusions section, the authors share their viewpoints on superconductivity and the SDGs.

## 2 Methodology

This review uses bibliometric analysis to investigate the scientific evolution of superconductivity and superconductors from 1951 to 2023, as well as their applications to achieve the Sustainable Development Goals (SDGs). We employed the bibliometric performance analysis technique and two data collections from Clarivate Analytics Web of Science (WoS) [64] to identify gaps and future research directions in this dynamic field. Database 1 was compiled from the WoS using the terms ("superconductivity" OR "superconductor") AND "application" in the titles, abstracts, and keywords of articles, covering the period from 1980 to 2023. We conducted a scientific mapping throughout this period, also analyzing the authors' keywords. Database 2 followed the same search method on the WoS database, focusing on the period from 2015 to 2023 and including the "Sustainable Development Goals" filter from the WoS database itself. This analysis aimed to understand the relationship between the field of superconductivity and its applications during the implementation of the UN SDGs (2015-2023). During the stages involving the analysis of the authors' keywords, a process was applied to correct typographical errors, plural forms, synonyms, and to eliminate articles without keywords. Figure 1 presents a flowchart detailing the methodology adopted in this research. The bibliometric analyses presented here were supported by various tools, including the Python and R programming languages, and the Bibliometrix package [65].

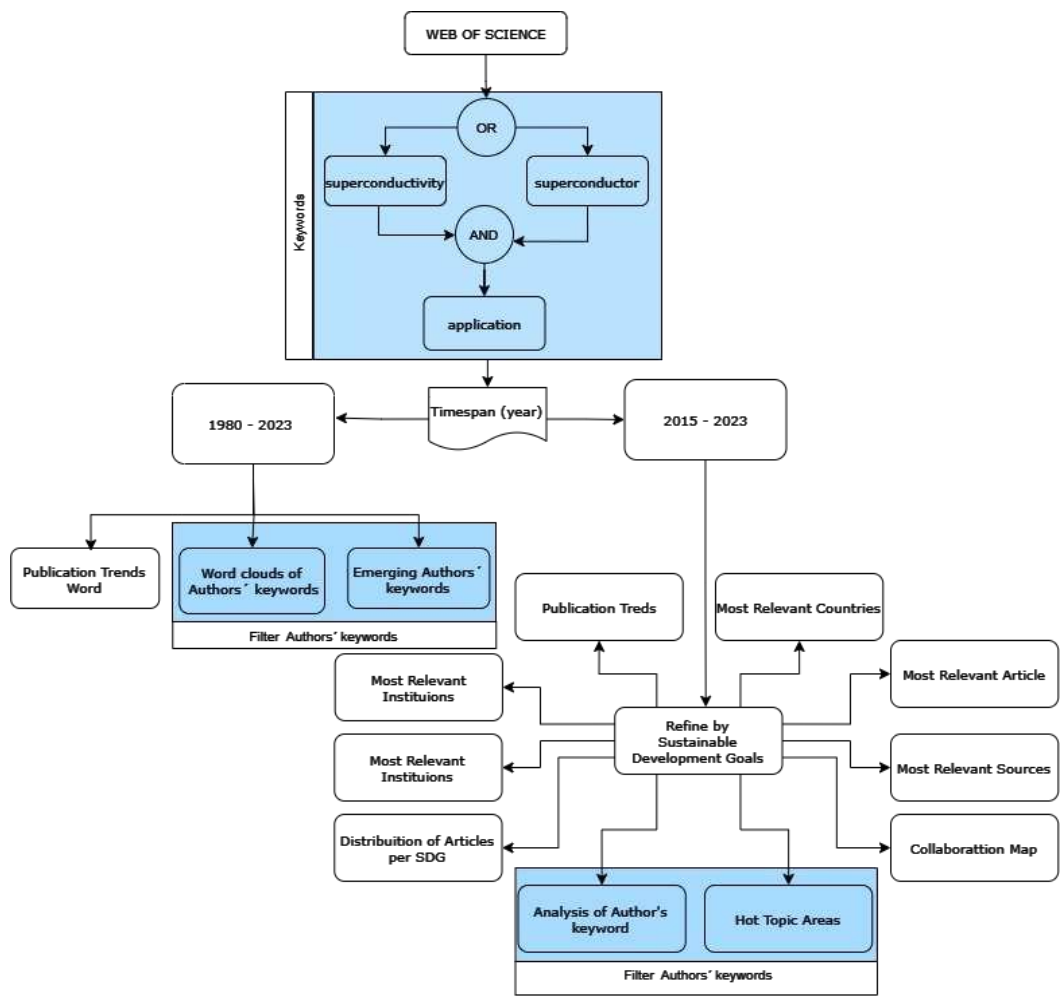

Figure 1: Research methodology flowchart for Data Collection and analysis.

## 3  A brief bibliometric analysis

This section examines the literature concerning the contribution of superconductivity and superconductors, as well as their applications for the SDGs. The data used in this section was collected from the Clarivate Analytics Web of Science (WoS) [64] with a total number of 19376 papers from 1980 to 2023. The subject term used was "superconductivity" OR "superconductor" AND "application". The covered document types are articles, proceedings papers, reviews, or meeting abstracts. In Figure 2, we can see that the volume of published research was negligible before 1986, i.e., the stage before the discovery of the first HTS, with no more than four (4) articles published per year, as seen in 1985. From 1986 to 1990, it is marked the onset of the HTS stage, the annual volume of published articles experienced modest growth, increasing from 3 in 1986 to 44 in 1990. Next, in the 1990s, the number of publications related to the applications of superconductivity and superconducting materials continued to increase, reaching its maximum annual growth rate in 1997 (551 articles), which was 186.1 times greater than that in 1986. Simultaneously, the 1990s were marked by turbulence, with the first negative effects of the economic development model, relying on unsustainable fossil fuels, well-documented in Refs. [66–69]. On one hand, novel and more sustainable development models emerged to mitigate economic and social inequalities [70–74]. On the other hand, within the environmental sphere, the United Nations Conference on Environment and Development (UNCED) occurred in 1992 at Rio de Janeiro city, Brazil [31]. This event catalyzed the establishment of various United Nations initiatives, including the Framework Con- vention on Climate Change (UNFCCC), which encompasses the Conference of the Parties (COP) [75, 76] and, more notably, the Kyoto Protocol at this stage [77]. Then, the annual growth rate of articles related to superconductivity and its applications in 2000 (410 articles) to 2001 (526 articles) was as high as 21.45%. At the same time, in 2001 the implementation of the UN's 2000 Agenda began through the implementation of the Millennium Development Goals (MDGs), initiating the MDGs stage, as illustrated in Figure 2. In general, the MDGs stage was characterized by a sustained increase in the number of articles per year, reaching a maximum of 754 articles in 2015. Subsequently, from 2015 to 2023, the annual growth of articles related to superconductivity and its applications reached unprecedented values. The yearly volume of published articles experienced the most significant growth, increasing from 754 in 2015 to 1026 in 2021. After implementing the UN's 2030 Agenda through its Sustainable Development Goals (SDGs) and the parties' binding nature of compliance with the Paris Agreement, there has been an unprecedented surge in publications on superconductivity and superconducting materials applications.

Researchers commonly use authors' keywords to describe their research's most notable content con- cisely. Hence, a detailed analysis of the authors' keywords could clarify and point out which are the consolidated research focus and hot topics in a research field. The word cloud, which contains the au- thors' keywords, will be one essential tool for presenting an overview of the current literature about superconductivity, sustainability, and SDGs [78]. To perform the authors' keywords analysis, we followed the strategy used by Pizzi et al. [79], in which, in the first step, we extracted the authors' keywords for each article in our dataset. These keywords were then filtered for duplicates, homogenized in spelling, and unique values were used to analyze our word cloud. The results of this analysis were presented in the word cloud of Figure 3. The font size represents the frequency of the authors' keywords. A word cloud analysis of the 150 most frequent keywords in the publication collection of Figure 2 was highlighted in Figure 3.

The 20 most frequent authors' keywords (see bigger keywords in the central area of Figure 3) were High-Temperature Superconductor (HTS, freq=1926), superconductor (freq=1404), magnetic (freq=946), superconductivity (freq=913), rebco (*RE*BCO, freq=873), critical current ($I_c$, freq=619), film (freq=514), windings or coil (freq=486), cable (freq=465), coated conductor (freq=458), josephson junction (freq=437), Magnetic Levitation (maglev, freq=401), magnesium diboride (MgB$_2$, freq=397), magnetic field (freq=360), alternating current loss (AC loss, freq=333), iron-based superconductor (freq=318), fault current limiter (freq=285), Superconducting Quantum Interference Device (SQUID, freq=272), modeling (freq=227), and Magnetic Resonance Imaging and Nuclear Magnetic Resonance (MRI & NMR, freq=222).

Excluding words intrinsically associated with superconductivity and superconducting materials (e.g., HTS, superconductor, critical current, superconductivity, and so on), the word cloud of Figure 3 can be seen from the keyword MgB$_2$ linked to the development of novel superconducting materials, which correspond to a consolidated research focus in our review. In addition, some authors' keywords associated with properties related to the phenomena of superconductivity and their applications, e.g., magnetic field, alternating current loss, fault current limiter appear as another consolidated research focus. Also, other authors' keywords explicitly linked to disruptive technological applications of superconductivity and superconducting materials such as windings or coil, cables, coated conductor, Josephson junction, maglev, AC loss, SQUID, MRI & NMR play a prominent role in the cloud of words.

Specific keywords emerged in the 2000s and showed an increasing trend through the SDGs stage, as shown in Figure 4. Rare earth barium copper oxide (*RE*BCO) and iron-based superconductor (IBS) key- words linked to HTS (or non-conventional superconductors) materials have recently increased frequency per year from 38 in 2015 to 70 in 2023 and 19 in 2015 to 42 in 2023, respectively. This increasing ap- pearance of iron-based superconductors can also be related to its almost recent discovery in 2008 [80–82]. Hence, research on *RE*BCO and iron-based superconductors emerged as hot topics in the SDG stage. This fact supports the idea that the discovery of novel HTS materials marked a technologically viable path for real applications of superconductivity, and the discovery of their non-conventional mechanisms. In addition, a similar trend can be extracted from hot topics related to the keywords windings or coil, cable, and Josephson junction. Recently, the most disruptive applications of superconductivity had an intense impact during the SDG stage on areas of sustainable human development, such as the development of green propulsion motors, efficient power transmission, scalable quantum information processing, and so on. It appears plausible to attempt establishing a correlation between the article growth rate related to superconductivity and its applications with the stage of the SDGs implementation depicted in Figure 2.

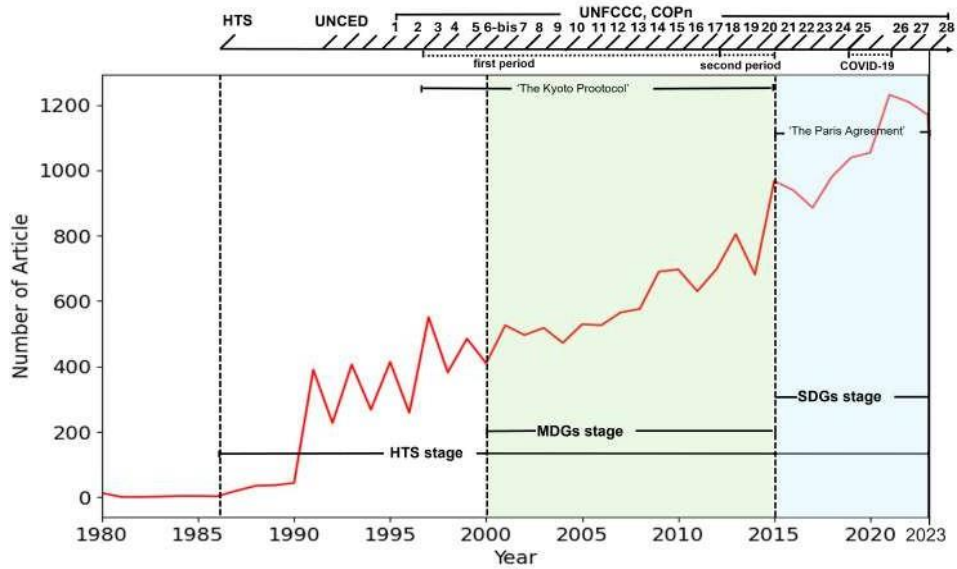

Figure 2: Number of publications per year related to superconductivity and superconductors and their applications from 1980 to 2023. This data collection is based on The Web of Science (WoS) [64], where the subject terms used were "superconductivity" OR "superconductor" AND "application".Abbreviations: High-Temperature Superconductors (HTSs), United Nations Conference on Environment and Develop- ment (UNCED), United Nations Framework Convention on Climate Change (UNFCCC), Conference of the Parties (COP), each COP is usually referred to by its number in the series, e.g., COP28 was the 28th COP meeting in Dubai, United Arab Emirates. Millennium Development Goals of the UN (MDGs) Sustainable Development Goals (SDGs.

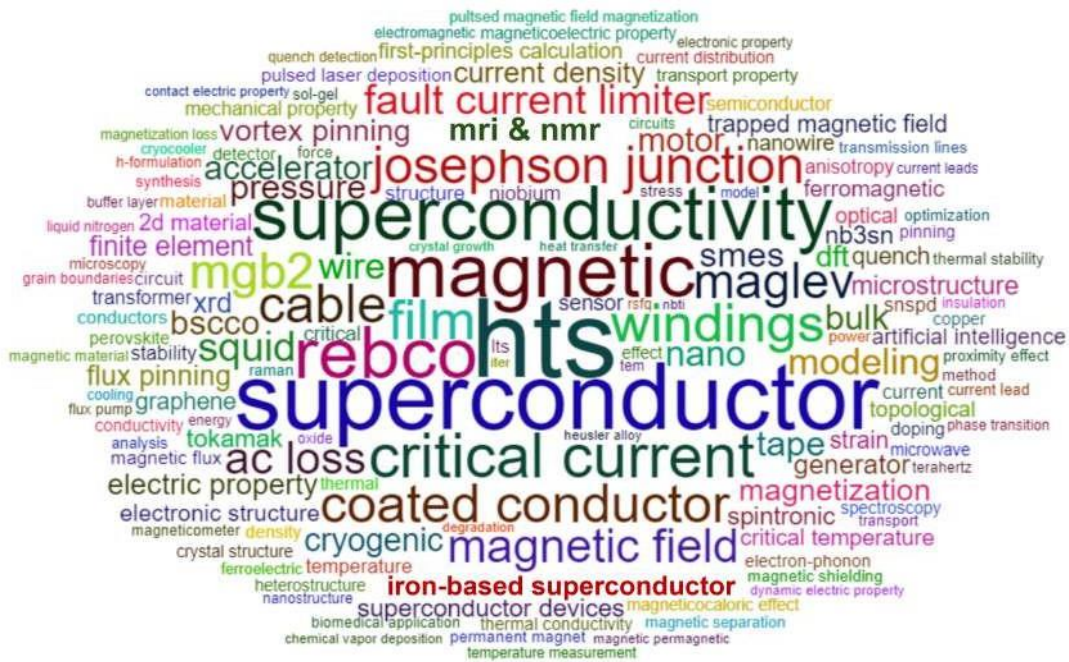

Figure 3: Word clouds of the 150 most frequent authors' keywords. From the publication collection of Figure 2

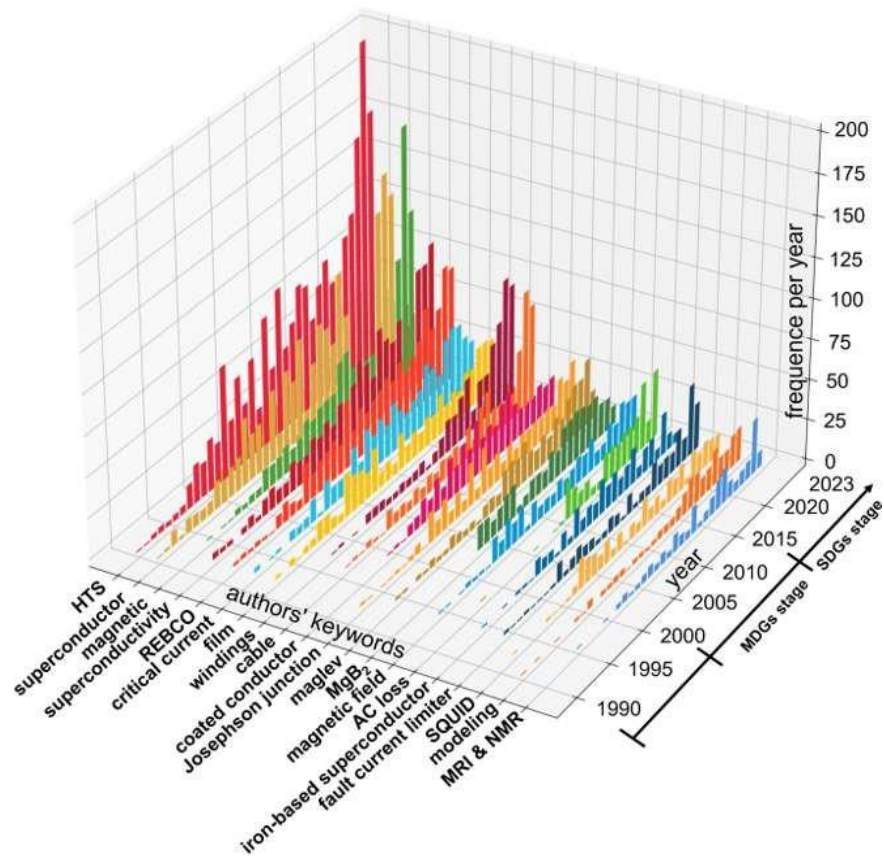

Figure 4: Evolution of frequences for some emerging keywords related to the superconductivity and superconductors and their applications from 1990 to 2023. From the publication collection of Figure 2.

# 4    Results of bibliometric analysis during the SDGs stage

## 4.1    Bibliometric indicators

To carry out this task, we deepen our study in the SDG stage refining the data by the Sustainable Development Goals filter of WoS, which uses a tagging and classification system based on the SDGs [83]. This search filter identifies which documents are directly related to the different SDGs. After our search using the subject term "superconductivity" OR "superconductor" AND "application", as well as the "Refine by Sustainable Development Goals" tool filter was completed, we obtained a total of 3164 papers from 2015 to 2023. Then, eliminating duplicities, we received 3053 papers from the WoS search, which covered articles, proceedings papers, reviews, or meeting abstracts.

Figures 5(a-b) shows the annual growth rate of publications, the cumulative number of articles (NA), and their citations concerning the applications of superconductivity, superconducting materials, and their links with the SDGs. Three thousand thirty-nine (3039) articles were cited 44620 times, and the average citations per year ranged between 0.8 in 2023 and 3.3 in 2020 during the SDGs stage (2015-2023). After the eighth year of the SDGs implementation, the number of articles published usually increased over the years. For example, the annual growth rate of articles in 2015 (295 articles) to 2022 (434 articles) was as high 47.1%. In addition, the evolution in the cumulative number of articles (CAN) from the start of the implementation stage to 2023 follows a second-order polynomial function of years variable (Y), CAN

$= B_2Y^2 + B_1Y + A$ (p $\leq$ 0.0001, $r^2$ = 0.9997), where $B_1$ = 8.03355 articles year$^{-2}$, $B_2$ = -32097.45736

articles year$^{-3}$, and A = 3.20627 × 10$^7$ articles year$^{-1}$. This equation allows us to build up a projection of the accumulative number of articles for the year 2030, i.e., the date on which the 2030 Agenda and its SDGs must be met. The predicted CAN value for 2030 was 6258, which could represent the most significant accumulation of knowledge in history in the field of superconductivity applications linked to sustainable development.

On the other hand, from 2015 to 2023, the 2015 year received the maximum number of citations (8487), and 2020 received the highest average citations per year (3.3) in the SDGs stage. These data emphasize the importance and impact of research related to the applications of superconductivity and superconductors in the field of sustainable UN's Goal development Agenda. Note that more recently published articles tend to have lower citation scores, as it takes some years for academics to cite the most recent research. In addition, next to 2021 there is a significant drop in citations, possibly as a consequence of COVID-19 disease declared as a global pandemic by the World Health Organization (WHO) in 2020 [84].

To assess the distribution of articles related to superconducting applications based on their classifica- tion by SDGs, we use the strategy according to Yamaguchi et al. [85], in which all articles were analyzed individually and classified according to the number of articles per year linked to a particular SDG, as shown in Figure 5(c). Quickly at the beginning of the implementation of the SDGs, a concern was ob- served in the research topics related to superconductivity and superconductors applied to achieve more affordable and clean energy, e.g., in 2015, 56.54% of articles were related to SDG 07, whereas 13.73% and 13.40% were related to SDG 03 (Good Health and Well-being) and SDG 09 (Industry, Innovation, and Infrastructure), respectively. In the range of 1-5%, SDGs 06, 11, and 13 were found, which are related to the goals of clean water and sanitation; sustainable cities and communities; and climate action. SDGs 01, 04, 08, and 10, linked to goals of social-economic sciences (see Table 1), were only found in the range of 0.5-1%. Additionally, in 2015, no articles linked to SDGs 02, 05, 15, 16, and 17 were published. At the onset of the SDGs stage, SDG 07 accounted for half of the publications. However, it wasn't until 2017 that the proportion of publications associated with SDG 07 decreased from 56.54% in 2015 to 47.54% in 2017. This decrease of approximately 9% for SDG 07 was offset by increases in the proportions for SDG 03 (Good Health and Well-being, 14.43%) and SDG 06 (Clean Water and Sanitation, 8.87%) in 2017. In 2022, the distribution of publications among the SDGs was the most embracing, with the fol- lowing percentages allocated to each SDG: SDG 01 (0.89%), SDG 02 (0.22%), SDG 03 (11.58%), SDG 04 (0.89%), SDG 06 (6.68%), SDG 07 (35.41%), SDG 08 (0.45%), SDG 09 (26.50%), SDG 10 (0.45%), SDG 11 (8.46%), SDG 13 (4.90%), and SDG 14 (1.34%). As expected, throughout the SDGs stage, applied and diversified research in the applications of superconductivity in the fields of energy, health, and the environment were observed. However, the results depicted in Figure 5(c) showed that articles in the superconductivity field were least linked to the SDGs concerning socioeconomic and human development. The SDGs are integrated, indivisible, balanced, and intersect across the three dimensions of sustainable development: economic, social, and environmental. Consequently, articles published within a certain field of research must have a balanced impact across all SDGs, demonstrating interdisciplinary and cross- cutting approaches to contribute to the 2030 Agenda effectively. We used the number of duplicate articles with the same DOI to calculate the percentage of interdisciplinary and cross-cutting articles published across different SDGs per year (% ICC$_{year}$) during the SDGs stage. It is evident that after 2018, there has been a noticeable increase in academic interest within the field of superconductivity, leading to the publication of articles with broader and cross-cutting implications for sustainable development (see % ICC$_{2018}$=1.79% and % ICC$_{2023}$ = 5.31% in Figure 5(c)). In 2020, the highest number of publications was recorded, while in 2022, the publications showed a higher average number of citations per year. Meanwhile, in 2023, the largest number of interdisciplinary and cross-cutting articles on the applications of superconductors and superconductivity to advance the SDGs were published (% ICC$_{2023}$ = 5.31%).

## 4.2    Most Productive Countries

Table 2 shows the twenty top corresponding authors' affiliated countries publishing papers on super-conductivity and its applications. China ranked first (NA = 886), followed by the United States (USA, NA = 440) and Japan (NA = 299). India and Germany had the fifth and sixth positions with 210 and 132 publications, respectively. Only the two countries with the largest and most powerful economies in the world (with values of GDP from 17.96 to 25.44$^{th}$ US\$) published more than 1300 articles from 2015 to 2023. The weight of China and the USA (NA = 1326 articles) over the total NA (NA = 3053 arti- cles) was 43.48%. The economic boom seems to encourage research related to superconductivity and its applications. In the ranks 2, 3, 5, 6, 8, and 10, six (6) of the intergovernmental association of the seven world's most advanced and developed economies (G7), namely: USA, Japan, Germany, Italy, United Kingdom (U.K), and France, concentrated 34.7% of articles in the SDGs stage. Yet, four of the five major emerging countries (BRICS): China (rank 1), India (rank 4), Russia (rank 9), and Brazil (rank 15), concentrated 39.8%. Therefore, publications by G7 and BRICS countries regarding the applications of superconductivity and SDGs were well-balanced in the studied period. Figure 6 shows the proportion of articles published on superconductivity and its applications per SDG for some G7 and BRICS member countries.

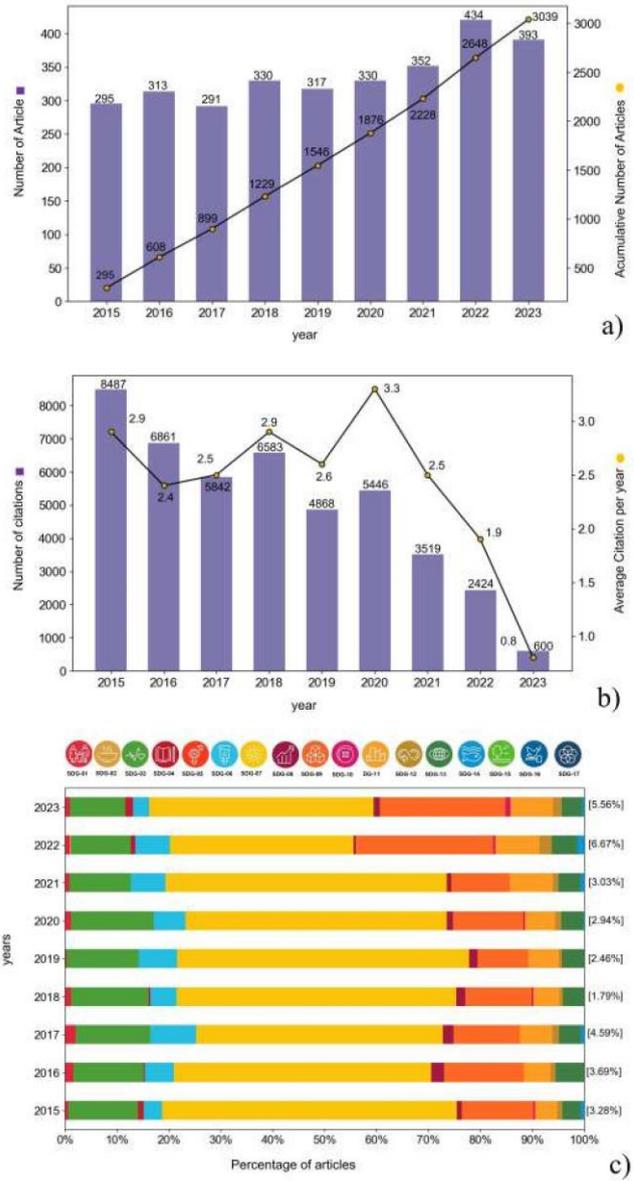

Figure 5: Number of articles per year, and cumulative scientific production without duplicity(a); number of citations per year, and average of citations per year (b); distribution of articles per SDG over the years of 2030 Agenda implementation.

Table 2: Top twenty most active countries based on number of articles per Corresponding Author's Countries

| Ranked based on Corresponding Author's Countries | | | | | | |
|---|---|---|---|---|---|---|
| Rank | Country | Continent | Association[a] | GDP (US$, in Tn)[f] | HDI[g] | NA |
| 1 | China | Asia | G20[b],BRICS[c] | 17.96 | 0.788 | 886 |
| 2 | United States (USA) | North America | G7[d], G20 | 25.44 | 0.927 | 440 |
| 3 | Japan | Asia | G7, G20 | 4.26 | 0.920 | 229 |
| 4 | India | Asia | BRICS, G20 | 3.42 | 0.644 | 210 |
| 5 | Germany | Europe | G7, G20, EU[e] | 4.08 | 0.950 | 132 |
| 6 | Italy | Europe | G7, G20, EU | 2.05 | 0.906 | 104 |
| 7 | Korea, Dem. People's Rep. | Asia | G20 | - | - | 103 |
| 8 | United Kingdom (U.K) | Europe | G7, G20 | 3.09 | 0.940 | 82 |
| 9 | Russia | Europe/Asia (Eurasia) | BRICS, G20 | 2.24 | 0.821 | 77 |
| 10 | France | Europe | G7, G20, EU | 2.78 | 0.910 | 67 |
| 11 | Switzerland | Europe | - | 0.81 | 0.967 | 55 |
| 12 | Pakistan | Asia | - | 0.37 | 0.540 | 48 |
| 13 | Algeria | Africa | - | 0.19 | 0.745 | 45 |
| 14 | Iran | Asia | - | 0.41 | 0.780 | 41 |
| 15 | Brazil | South America | BRICS, G20 | 1.92 | 0.760 | 37 |
| 16 | Australia | Oceania | G20 | 1.69 | 0.946 | 35 |
| 17 | Turkey | Europe/Asia | G20 | 0.90 | 0.855 | 32 |
| 18 | Poland | Europe | EU | 0.69 | 0.881 | 31 |
| 19 | Bangladesh | Asia | - | 0.46 | 0.670 | 26 |
| 20 | Egypt | Africa | - | 0.48 | 0.728 | 25 |

[a] Intergovernmental Association. [b] "Group of Twenty" (G20) − An intergovernmental association that brings together major advanced and emerging economies. [c] BRICS − An acronym for the intergovernmental association of five major emerging countries: Brazil, Russia, India, China, and South Africa. [86] [d] Gross Domestic Product (GDP). [e] Intergovernmental asso- ciation of the world's seven most advanced and developed economies (G7). [f] European Union (EU). [g] Human Development Index (HDI)[87]

The most industrialized countries in the world (i.e., G7) published more articles related to super-conductivity and its applications with a marked link to SDGs 03, 07, and 09 that address central issues related to health, clean energy, and sustainable industries, respectively. Thus, the percentage of publica- tions linked to SDGs 03, 07, and 09 for the USA, Japan, Germany, Italy, and the U.K. reached values of 92.19%, 85.09%, 93.33%, 92.31%, and 85.08%, respectively. On the contrary, the G7 articles had little influence on SDG 13 and 14, seeking actions to remedy climate change and restore the global balance of life on earth (≤ 3%). Conversely to G7 countries, BRICS countries seem to publish more influential articles on SDGs 13 and 14, with the percentage of articles linked to SDGs 13 and 14 of 4.53%, 14.92%, 14.1%, and 9.09% for China, India, Russia, and Brazil, respectively. The USA was the most influential country for SDG 03, whereas China was the most influential for SDG 07. Interestingly, India presented the most prominent role in articles linked to the goal of "Clean water and sanitation" (SDG 06) with 25.13%. Additionally, India and Japan had the highest number of SDGs (up to 13) among all the top 20 countries studied from 2015 to 2023. Finally, Brazil's distribution of articles for SDGs 03, 07, and 09 (73.0%) was closer to the G7 countries than to the BRICS ones.

On the other hand, some countries that are part of the European Union and the G20, e.g., Australia, Turkey, and Poland, barely accounted for 2.7% of articles in the SDGs stage. Curiously, some countries with a medium or low human development index (HDI ≤ 0.760) appear in the Top 20 influential countries in research focused on superconductivity and its applications, e.g., Pakistan (rank 12), Algeria (rank 13), Bangladesh (rank 19), and Egypt (rank 20).

Robust collaboration among international institutions facilitates extensive superconductivity and SDG research. From a geographical point of view, all continents were represented in the Top 20 influential countries. Furthermore, Figure 7 shows a country collaboration map. This map indicates the collabo- rative networks by straight lines connecting two or more countries. Notably, China and the USA were leaders in collaboration, followed by Japan and Germany. Excluding China, the leading global collab- oration networks were mainly between G7 countries. Australia (G20) and Switzerland are prominent in collaborations between the USA, Germany, and Japan. Although the publication of articles for both the G7 and the BRICS was balanced in the SDGs stage, the G7 countries achieved much more solid collaboration networks in the period studied. Consequently, emerging or developing countries have the challenge of completing the consolidation of networks of collaborators at a global level. This fact could positively impact the UN 2030 Agenda.

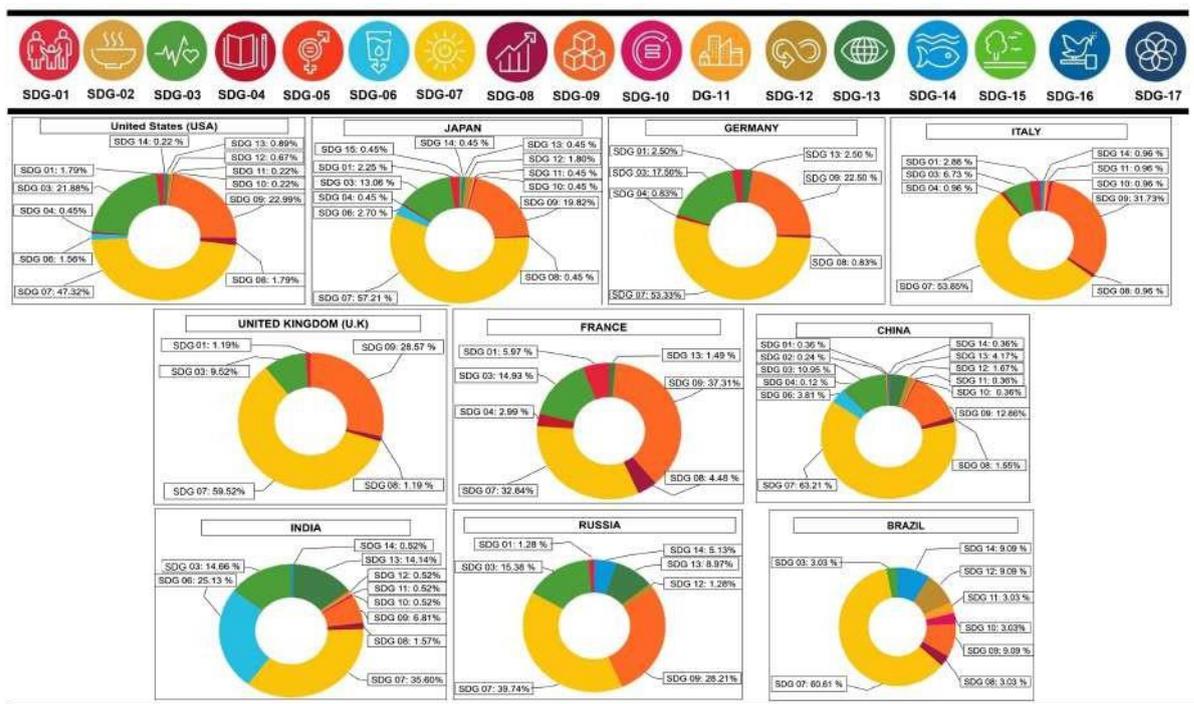

Figure 6: Distribution of articles per SDG for some G7 and BRICS member countries from 2015 to 2023.

## 4.3   Most Relevant Institutions/Organizations

During the SDGs stage, 1,760 institutions contributed to 3,053 publications related to superconduc- tivity and its applications. Achieving the UN's 2030 agenda is largely dependent on the efforts of the governments of UN member countries. However, reaching the SDGs and fulfilling the 2030 agenda also requires substantial contributions from the leading research institutions and universities globally. Table 3 lists the Top 10 most productive institutions/organizations by the number of articles on superconductivity and its applications linked to the SDGs.

Table 3: Top 10 productive institutions/organizations in research on superconductivity and its applica- tions

| | | Ranked by NA | | | | |
|---|---|---|---|---|---|---|
| Rank | Organization/ Institution | ref.[a] | Country | Association | NA[b] | TC[c] | CA[d] |
| 1 | United States Department of Energy | [88] | USA | G7, G20 | 246 | 4599 | 18.7 |
| 2 | University of Houston System | [89] | USA | G7, G20 | 177 | 7064 | 39.9 |
| 3 | University of Chinese Academy of Sciences, CAS[e] | [90] | China | BRICSc, G20 | 133 | 1895 | 14.2 |
| 4 | Nanjing University | [91] | China | BRICS, G20 | 127 | 1653 | 13.0 |
| 5 | Centre National de la Recherche Scientifique | [92] | France | G7, G20, EU | 108 | 2388 | 22.1 |
| 6 | Helmholtz Association | [93] | Germany | G7, G20, EU | 105 | 2802 | 26.7 |
| 7 | University of California System | [94] | USA | G7, G20 | 92 | 2874 | 31.2 |
| 8 | University of Tokyo | [95] | Japan | G7, G20 | 89 | 1557 | 17.5 |
| 9 | Institute of Physics, CAS | [96] | China | BRICS, G20 | 82 | 1229 | 15 |
| 10 | National Institute for Materials Science | [97] | Japan | G7, G20 | 77 | 968 | 12.6 |

[a] Website. [b] Number of articles, CA. [c] total citations, CT. [d] Citation per article, CA. [e] Organization/Institution linked to the (CAS) [98]

The United States Department of Energy (DOE) leads with 246 articles and 4,599 citations, followed by the University of Houston System, which ranks second with 177 articles and 7,064 citations. The University of Chinese Academy of Sciences is ranked third, with 133 articles and 1,895 citations. Notably, at least two institutions of the Top 3 encompass multiple locations and complex organizations. In contrast, the University of Chinese Academy of Sciences is part of the larger conglomerate of institutes within

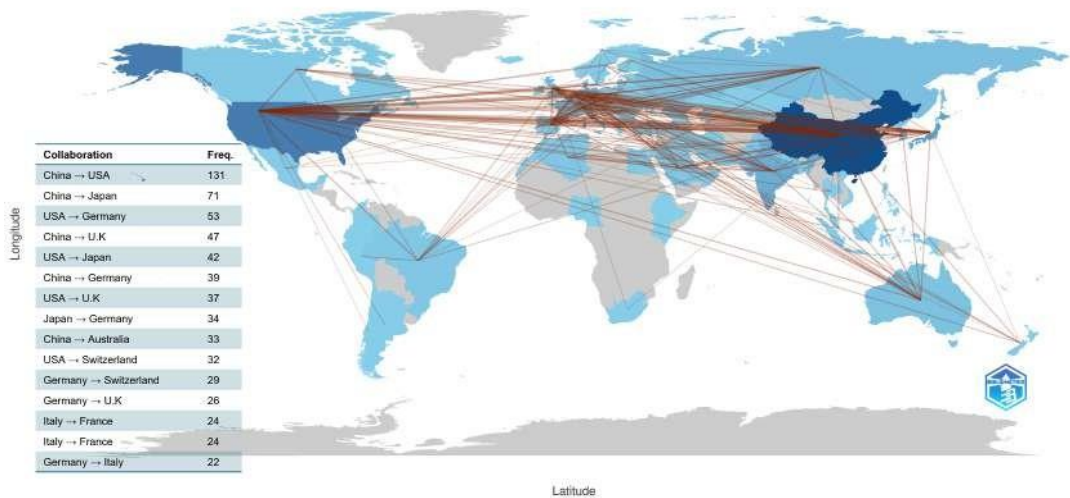

Country Collaboration Map

| Collaboration | Freq. |
|---|---|
| China → USA | 131 |
| China → Japan | 71 |
| USA → Germany | 53 |
| China → U.K | 47 |
| USA → Japan | 42 |
| China → Germany | 39 |
| USA → U.K | 37 |
| Japan → Germany | 34 |
| China → Australia | 33 |
| USA → Switzerland | 32 |
| Germany → Switzerland | 29 |
| Germany → U.K | 26 |
| Italy → France | 24 |
| Italy → France | 24 |
| Germany → Italy | 22 |

Figure 7: Country collaboration Map for research on superconductivity and its applications in the SDGs stage.

the Chinese Academy of Sciences (CAS). Due to the complexity of CAS, Chinese scholars often use varied syntax in their affiliations, complicating the calculation of the total number of articles attributable to CAS. Nevertheless, the Institute of Physics, affiliated with CAS, ranks ninth with 82 articles and 1,229 citations during the SDGs stage. Here, we only briefly summarize this fact, but a more in-depth analysis of the Chinese case is extensively described by Zhu et al. [99]. The majority of these ten institutions were from G7 countries, including the USA (3), Japan (2), Germany (1), and France (1). Meanwhile, China, with three (3) institutions, represented the BRICS countries. Notably, developing countries did not have influential institutions during the SDGs period. Regarding citations per article, the University of Houston System received the highest score (CA = 39.0), followed by the University of California System (CA = 31.2) and the Helmholtz Association (CA = 26.7). The formation of networks among multidisciplinary institutions clearly generated research with greater impact and influence among superconductivity scholars during the SDG stage. This approach could serve as a guideline for developing countries, including BRICS and G20 members, to build more successful academic systems in the field of superconductivity and to advance the SDGs.

## 4.4 Most Relevant Sources

Table 4 presents the Top ten (10) most prolific journals from the research area under study. The weight of the most productive journals according to the number of articles (NA = 1,554 articles) over total NA in the SDGs stage was 4.46%, whereas the Top three (3) represented 3.18% over total NA. This weight was significant considering that the Top 10 of the most productive journals comprise only 1.7% of the total number of specialized journals in the review (577 journals). Among the Top four (4) stand out: *IEEE Transactions on Applied Superconductivity* (423 articles), *Journal of Superconductivity and Novel Magnetism* (404 articles), *Superconductor Science & Technology* (211 articles), and *Physical Review B* (169 articles). The scope and goals of these four specialized journals were broad enough to collect the most significant number of publications related to superconductivity and its applications from 2015 to 2023. Figure 9 highlights the relationship between the distribution of articles published in the Top 10 journals (see the left section in Table 4) focusing on superconducting applications and their respective SDGs.

It is worth noting that articles published in the Top 10 journals during the SDGs stage had a greater impact in the fields of superconductivity and clean energy, with percentages of articles linked to SDG 07 (Affordable and clean energy) ranging from 22.25% to 90.91%. The articles of the journal's Top 10 had a balanced impact on the field of health and its SDG 03, with publication weights in the range of 3.23- 16.20%. Conversely, the ranks from 7 to 10 in the Top 10 only presented links between their publications and SDG 6 smaller than 3.03%. In addition, the articles published in the *Journal of Superconductivity and Novel Magnetism* (see rank 2) intensely influenced SDG 06 during its implementation stage. Lastly, a more comprehensive range of colors in Figure 9 signifies a more substantial global alignment with the UN 2030 Agenda. Consequently, the journals ranked between positions 4 and 7 exhibited the highest influence on the UN 2030 agenda, demonstrating a significant impact on SDGs 01, 03, 06-09, and 11-13. The most influential journals regarding the total citations are presented in the central section of Table 4. Similar to the ranking by the number of articles, the top four ranks in terms of total citations were held by the journals *Journal of Superconductivity and Novel Magnetism*, *Superconductor Science & Technology*, *IEEE Transactions on Applied Superconductivity*, and *Physical Review*, collectively representing over 40% of the Top 10 according to citations. When considering the citation per article to a new ranking (see the right section in Table 3), *Nature Reviews Materials* emerges as the clear leader with an impressive average of 302.3 citations per article. In addition to the Top 3, *Materials Today* and *Journal of Optics* presented 285.0 and 265.0 citations per article, respectively.

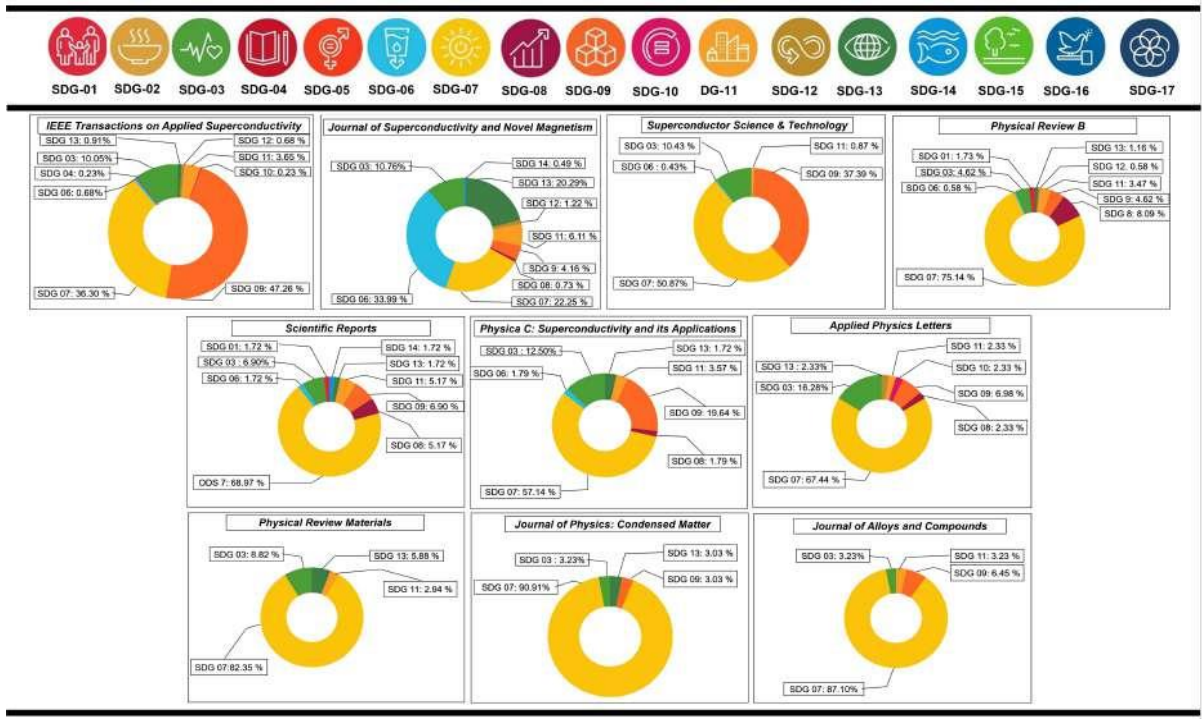

Figure 8: Distribution of articles per SDG and institutions/organizations from 2015 to 2023.

While citation analysis (CA) values indicate the influence of articles published in a journal regarding superconductivity and Sustainable Development Goals (SDGs), the impact factor (IF) values provide insight into the general impact of articles published in a journal. With the exception of the *Journal of Optics* and *Frontiers in Chemistry*, CA values are closely related to the respective IF of each journal. Unfortunately, the three journals with the highest impact only focus on five (5) articles related to the applications of superconductivity and SDGs between 2015 and 2023. Academics in the field of supercon- ductivity should prioritize journals with higher CA and IF, as this could lead to the creation of more diverse graphics, such as the one presented in Figure 9 (e.g., see percentages of articles by SDG for the *Physical Review B* in Figure 6), and consequently, a more significant global impact on the 2030 Agenda.

## 4.5 Most Relevant Articles

Between 2015 and 2023, i.e., the SDGs stage (see Figure 2), numerous articles were published on the applications of superconducting materials. Recognizing highly cited articles opens a window to delve deeper into the state of the art in any field of knowledge. So, Table 5 presents the Top ten (10) most cited articles linked to SDGs from the research area under study. The weight of these Top 10 papers over the total number of citations in the SDGs stage was equal to 10.07%, whereas the Top 3 most globally cited articles represented 4.62%. This fact showed that top research significantly impacts citation analysis on the area of superconductivity.

Additionally, the Top 10 are highly multidisciplinary, e.g., the articles address topics ranging from theoretical aspects and computational modeling of superconductivity to the synthesis of superconduct-ing materials for diverse applications such as photocatalysis and photosensors. The article "Large-area high-quality 2D ultrathin $Mo_2C$ superconducting crystals" by Xu et al. [101] ranked first, with 946 ci- tations and a total citation per year of 95. This article reported a transition metal carbide known for its superconductivity and robust thermal and chemical stability. The research achieved defect-free 2D superconductivity, representing a superconductor in the clean limit. By linking to WoS's Sustainable Development Goals (SDGs), a direct association between the article and SDG 11 (Sustainable cities and communities) was established. In addition, our in-depth analysis of the Xu et al. [101]'s article indicates that these findings hold promise for applications in superconducting devices and advancing fundamen- tal science, aligning with SDG 7 (Affordable and clean energy) and SDG 9 (Industry, innovation, and infrastructure).

Table 4: Top ten most active journals based on the number of articles, and its citations.

| | Ranked according to NA | | Ranked according to Tc | | Ranked according to CA | | |
|---|---|---|---|---|---|---|---|
| Rank | Journal[a] | NA[b] | Journal | TC[c] | Journal | CA[d] | IF[e] |
| 1 | IEEE Trans. Appl. Supercond. | 423 | J. Supercond. Novel Magn. | 3588 | Nat. Rev. Mater. | 302.3 | 83.5 |
| 2 | J. Supercond. Novel Magn. | 404 | Supercond. Sci. Technol. | 2827 | Mater. Today | 285.0 | 24.2 |
| 3 | Supercond. Sci. Technol. | 211 | IEEE Trans. Appl. Supercond. | 2358 | J. Opt. | 265.0 | 2.1 |
| 4 | Phys. Rev. B | 169 | Phys. Rev. B | 2158 | Nat. Mater. | 252.8 | 41.2 |
| 5 | Sci. Rep. | 55 | Nanoscale | 1508 | Front. Chem. | 238.0 | 5.5 |
| 6 | Physica C | 53 | Rev. Mod. Phys. | 1310 | Rev. Mod. Phys. | 218.3 | 44.0 |
| 7 | Appl. Phys. Lett. | 41 | Nat. Mater | 1264 | Phys. Rep.-Rev. Sec. Phys. Lett. | 166.0 | 29.9 |
| 8 | Phys. Rev. Mater. | 34 | Nat. Phys. | 1130 | Adv. Energy Mater. | 154.5 | 27.8 |
| 9 | J. Phys. Condens. Matter. | 33 | Sci. Rep. | 1104 | Nanoscale | 150.8 | 6.7 |
| 10 | J. Alloys Compd. | 31 | Nat. Rev. Mater. | 907 | Science | 146.3 | 56.9 |

[a] Abbreviation according to ISO4, IEEE Transactions on Applied Superconductivity: IEEE Trans. Appl. Supercond.; Journal of Superconductivity and Novel Magnetism: J. Supercond. Novel Magn.; Superconductor Science & Technology: Supercond. Sci. Technol., 4 Physical Review B: Phys. Rev. B; Scientific Reports: Sci. Rep.; Physica C: Superconductivity and its Applications: Physica C Supercond. Appl.; Applied Physics Letters: Appl. Phys. Lett.; Physical Review Materials: Phys. Rev. Mater.; Journal of Physics: Condensed Matter: J. Phys. Condens. Matter.; Journal of Alloys and Compounds: J. Alloys Compd.; Reviews of Modern Physics: Rev. Mod. Phys.; Nature Materials: Nat. Mater.; Nature Physics: Nat. Phys.; Nature Reviews Materials Nat. Rev. Mater.; Materials Today: Mater. Today; Journal of Optics: J. Opt.; Frontiers in Chemistry: Front. Chem.; Physics Reports (Review Section of Physics Letters): Phys. Rep. Phys. Lett.; Advanced Energy Materials: Adv. Energy Mater.. [b] Number of articles, CA. [c]total citations, CT. [d] Citation per article, CA. [e] IF, Impact factor according to Journal Citation Reports [100].

Interestingly, the articles published by Mcardle et al. [102] and Jain et al. [103], which employ theoretical approaches to both understand superconductivity and how superconductors could contribute to the quantum computing field, were ranked second and third, with 946 and 620 citations, respectively. These articles exemplify a diverse range of theoretical and computational tools used in the study and application of superconductivity. According to WoS's classification, these articles are linked to SDG 9 (Industry, innovation, and infrastructure) and SDG 07, respectively.

From Table 5, it becomes clear that the most globally cited articles have a broad and significant impact, spanning several SDGs, notably SDG 03 (Good health and well-being), SDG 06 (Clean water and sanitation), SDG 07 (Affordable and clean energy), SDG 09 (Industry, innovation, and infrastructure), and curiously, SDG 11 (Sustainable cities and communities). On one hand, as expected, the most cited articles were published in influential journals. On the other hand, the majority of the scientific community publishes their applied works in journals that do not appear in the Top 10 according to NA.

# 5  Dynamics of Keywords and trend topics during the SDGs stage

## 5.1  Most Relevant subject Areas

Table 6 shows the Top 10 subject areas classified by WoS thematic categories on keywords superconductivity, superconductors, and applications contributing to SDGs based on the number of articles. The results showed that the five main areas covered based on the number of articles were (rank 1) Physics, (rank 2) Materials Science, (rank 3) Engineering, (rank 4) Chemistry, and (rank 5) Science & Technology- Other Topics. Meanwhile, the more interdisciplinary areas, such as Instruments & Instrumentation (rank 8), Energy & Fuels (rank 9), and Thermodynamics (rank 10), concentrated the smallest number of articles in the SDGs stage. Note that when considering the total citations and citations per article as an area impact factor, the areas linked to Science & Technology, Energy & Fuels, and chemicals concentrated the greatest attention for superconductivity scholars in the SDGs stage. The findings of this research will be helpful for researchers and policymakers to understand the critical subject areas contributing to superconductivity, superconductors, and their applications linked to Sustainable Development Goals and 2030 Agenda.

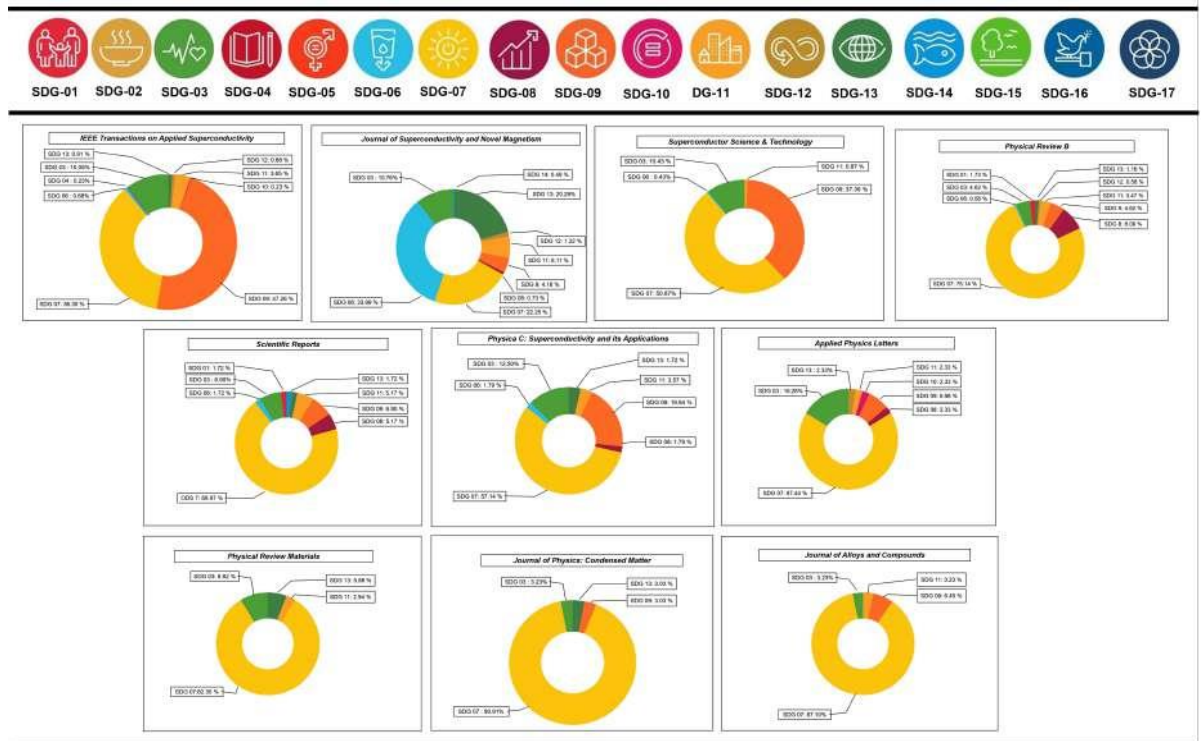

Figure 9: Percentage of articles per SDG for the Top ten (10) most prolific journals

## 5.2    Authors' keywords analysis, Thematic Focus, and Hot Topic

We performed a deeper analysis of the occurrences (i.e., frequency) and co-occurrences (i.e., co- frequency) of authors' keywords, focusing on 150 keywords declared by 150 superconductivity researchers. These keywords appeared at least ten times in the collection, and four times during the data collection stage for SDGs implementation (see Figure 10). To ensure consistency, we filtered out duplicates, stan- dardized spellings, and excluded irrelevant keywords (e.g., superconductor, HTS superconductivity, etc) from the analysis. The size of each circle (node) represents the frequency of each author's keyword, while the thickness of the lines between nodes indicates the strength of their co-occurrence, based on how often they appeared together in published articles. This analysis revealed that research related to supercon- ductivity, superconductors, and their applications during the SDG stage could be divided into thirteen (13) clusters in the co-keyword network (Figure 10). The general characteristics of these clusters are described in Table 7. Additionally, we examined up to the Top 10 keywords in each cluster to determine their thematic core and linkage to SDGs, calculating the percentage of keywords associated with each SDG (see Figure 11).

The biggest cluster (cluster 1, green color, ●) around the keywords iron-based, Niobium and $MgB_2$ presented the top ten most frequent author's keywords. The keywords iron-based, $MgB_2$, film, wire, Josephson junction, tape, vortex pinning, strain, maglev, Niobium, appeared with frequencies of 154, 146, 67, 53, 39, 35, 30, 27, 24, and 24 times, respectively. The green cluster is mainly associated with the thematic core on traffic, sensor, and quantum device applications. Notable applications include the production of propulsion motors and Superconducting Nanowire Single Photon Detectors (SNSPDs) using both $MgB_2$ and NbN nanowires. The highest values of average frequency (AF) and its standard deviation (SFAF, 24.81 ± 32.14) indicate a high degree of scope and dispersion among the nodes in the green cluster.

Cluster 2 (purple color, ●), shares the same number of keywords (NK = 38) as Cluster 1. The most popular keywords in Cluster 2 are windings (frequency = 146), cable (106), MRI & NMR (70), *RE*BCO (69), SMES (50), Nb$_3$Sn (49), tokamak (42), cryogenic (41), electric properties (35), and motor (35). Although Cluster 2 has slightly lower frequency values (20.08 $\pm$ 28.59) than Cluster 1, it occupies a more central location in the keyword network (see Figure 10). Consequently, it shows higher co-frequency and a greater degree of connections with other clusters in the network, forming a more general and transversal thematic core. Cluster 2 is mainly linked to the health and energy thematic core, e.g., some superconducting materials in the form of tapes, bulk, and cables, are employed in power generators. In addition, the superconductor plays a fundamental role in possible new magnets based on *RE*BCO for different MRI & NMR equipment.

Table 5: Top 10 most global cited articles.

| Rank | Authors[a] | Title | TC[b] | Journal | Year | SDG |
|------|-----------|-------|------|---------|------|-----|
| 1 | Xu *et al.* [101] | Large-area high-quality 2D ultrathin $Mo_2C$ superconducting crystals | 946 | *Nat. Mater.* | 2015 | SDG 11 |
| 2 | Mcardle *et al.* [102] | Quantum computational chemistry | 620 | *Rev. Mod. Phys.* | 2020 | SDG 9 |
| 3 | Jain *et al.* [103] | Computational predictions of energy materials using density functional theory | 499 | *Nat. Rev. Mater.* | 2016 | SDG 7 |
| 4 | Miransky *et al.* [104] | Quantum field theory in a magnetic field: From quantum chromodynamics to graphene and Dirac semimetals | 450 | *Phys. Rep.* | 2015 | SDG 11 |
| 5 | Smejkal *et al.* [105] | Topological antiferromagnetic spintronics | 392 | *Nat. Phys.* | 2018 | SDG 11 |
| 6 | Casola *et al.* [106] | Probing condensed matter physics with magnetometry based on nitrogen-vacancy cen- tres in diamond | 346 | *Nat. Rev. Mater.* | 2018 | SDG 3 |
| 7 | Yang *et al.* [107] | A Mini Review on Carbon Quantum Dots: Preparation, Proper- ties, and Electrocatalytic Appli- cation | 324 | *Adv. Mater.* | 2017 | SDG 11 |
| 8 | Hu *et al.* [108] | Topological polaritons and photonic magic angles in twisted $\alpha$-$MoO_3$ bilayers | 314 | *Nature* | 2020 | SDG 7 |
| 9 | Neu *et al.* [109] | Tutorial: An introduction to ter- ahertz time domain spectroscopy (THz-TDS) | 305 | *J. Appl. Phys.* | 2018 | SDG 3 |
| 10 | Xue *et al.* [110] | Opening Two-Dimensional Materials for Energy Conversion and Storage: A Concept | 301 | *Adv. Energy Mater.* | 2017 | SDG 11 |

[a] In this section, data from the article by Zhao et al. [111] in Nanoscale was discarded despite its link to SDG 11. Although it mentions superconductivity, it needs more relevant content regarding its application within the superconductivity field and SDGs and, therefore, was excluded from the discussion.
[b] Abbreviation of total citation

Table 6: Subject area according to WoS thematic categories in research on superconductivity and its applications linked to SDGs

| Rank | Subject Area | NA[b] | TC[c] | CA[d] |
|------|-------------|------|------|------|
| 1 | Physics | 2242 | 29653 | 13.2 |
| 2 | Materials Science | 823 | 16643 | 20.2 |
| 3 | Engineering | 636 | 4064 | 6.4 |
| 4 | Chemistry | 427 | 11939 | 28.0 |
| 5 | Science & Technology-Other Topics | 373 | 11617 | 31.1 |
| 6 | Metallurgy & Metallurgical Engineering | 72 | 663 | 9.2 |
| 7 | Optics | 67 | 998 | 14.9 |
| 8 | Instruments & Instrumentation | 51 | 485 | 9.5 |
| 9 | Energy & Fuels | 48 | 1406 | 29.3 |
| 10 | Thermodynamics | 44 | 283 | 6.4 |

[b] Number of articles, NA. [c] Total citations, TC. [d] Citation per article, CA.

Interestingly, Cluster 3 (in orange, ●) appears in the lower zone of the keyword network. This cluster is related to the thematic core of theoretical and computational modeling of structural materials and their superconductivity. The main characteristic of Cluster 3 is its co-frequency solid linkage with both the largest clusters (Cluster 1 and Cluster 2) and some smaller clusters related to the thematic core of developing novel materials with disruptive superconducting nano-scale applications (Cluster 4, pink color, ●) and micro-scale applications (Cluster 7, brown color,●). In addition, Cluster 5 (in light green, ●) emerges, concentrating on keywords related to the experimental characterization of superconducting materials. This cluster also shows a strong co-frequency link with the other clusters in the network. Finally, the miscellaneous clusters (see Clusters 6, 8-13) located on the periphery of the keyword network condense thematic cores associated with the general properties of superconductivity, superconductors, and their applications. Figure 12 provides an overview of the most highlighted trend topics during the SDG implementation stage. The size of each circle in the chart corresponds to the frequency of each author's keyword or hot topic researched by superconductivity academics in a given year. The length of the horizontal lines around these hot topics indicates the longevity of the research field; longer lines represent more consolidated research focus. Early when the SDGs stage began (from 2015 to 2017), it was noted how some applications of some magnetic material based on superconductors appeared among the research topics of superconductivity academics. Then, between 2018 and 2020, thematic cores related to the health area emerged (e.g., applications in MRI & NMR to clinic diagnostic technologies), and disruptive sustainable energies (e.g., coils to Tokamak technologies). The trend topics MRI & NMR and modern particle accelerators were the consolidated research focus from 2018 to 2020. In addition, the keywords windings, cables, and SEMES include some sustainable core thematic, such as green propulsion motors, efficient power transmission, and Superconducting Magnetic Energy Storage (SEMES). It seems that as progress is made in the timeline of the SDGs (between 2021 and 2023), superconductivity academics are more concerned about topics related to sustainable development.

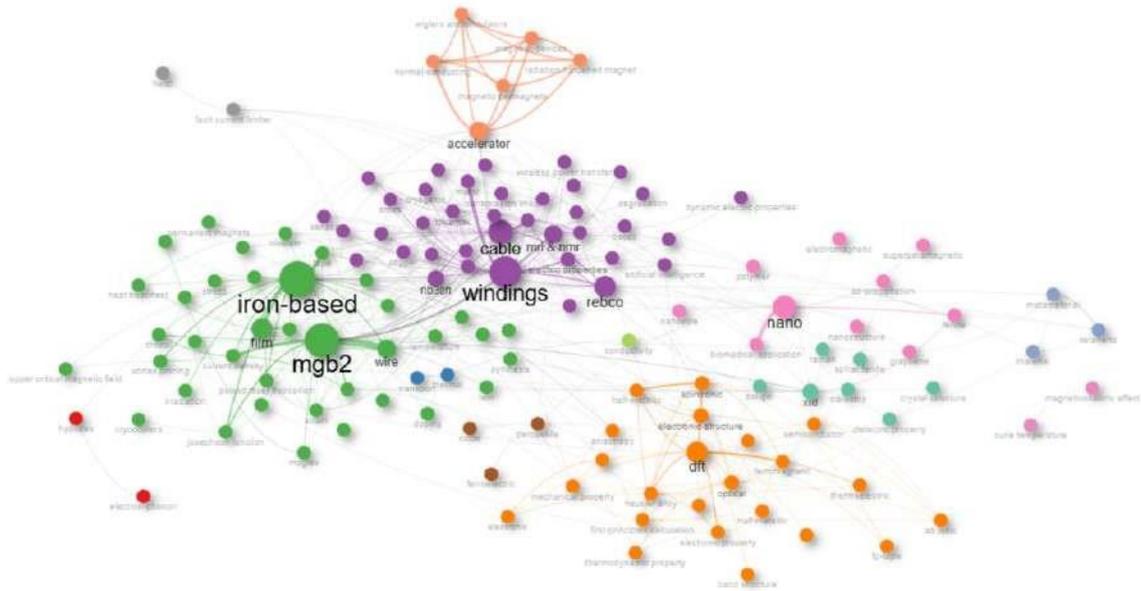

Figure 10: Clusters of authors' keywords based on co-occurrence of the keywords with at least 5 occur- rences (150 keywords), map created with bibliometrix package [65] on the SDGs stage.

The keywords of the green biggest cluster are mainly linked to SDGs 07 (66.25%), SDG 09 (20.55%), and SDG 03 (6.71%), and to a lesser extent (¡ 5%) to SDGs 01, 02, 06, and 11-13 (see Figure 10). Similarly, the purple central cluster (Cluster 2's keywords) mainly linked to SDGs 09 (56.32%), SDG 07 (22.29%), and SDG 03 (9.66%), and to a lesser extent (¡ 5%) to the SDGs 01, 04, 06, and 11-14. The distribution of keywords among the SDGs was more embracing in Cluster 3, with the following percentages: 46.2% (SDG 07), 33.54% (SDG 13), 6.96% (SDG 06), 5.7% (SDG 03), 3.16% (SDG 09), 2.85% (SDG 11), and 1.58% (SDG 12).

Regardless of the thematic core of the clusters in the keyword network of Figure 10, most author keywords are further linked to the Sustainable Development Goals (see Figure 11). For example, the development of new magnets for NMR with lower energy consumption and production costs could be  a primary goal aligned with the SDGs. In particular, the superconductivity link to medical diagnosis contributes to SDG 03 (i.e., promotion of health and well-being), enhancing procedure accuracy, reducing the need for invasive interventions, and consequently improving the life expectancy of patients. Addi- tionally, more accurate clinical diagnoses can contribute to reducing health inequalities (i.e., SDG 10), ensuring that everyone has access to advanced medical care, as well as for SDG 09 in terms of innovation, infrastructure and industry. Therefore, superconductivity plays a significant role in pursuing a healthier, more sustainable future that is aligned with global development goals. Additionally, developing more efficient and environmentally friendly aircraft turbines contributes to SDG 13 (Climate Action) and SDG 09 (Industry, Innovation, and Infrastructure). By reducing fuel consumption and polluting gas emissions,

Table 7: General characteristics of keywords clusters.

| Cluster | Top Keywords per Cluster | $NK^a$ | $AF^a \pm SDAF^a$ | Linked SDG |
|---|---|---|---|---|
| 1 (●) | Iron-based (154), MgB$_2$ (146), film (67), wire (53), Josephson junction (39), tape (35), vortex pinning (30), strain (27), maglev (24), Niobium (24). | 38 | 24.81 ± 32.14 | 01, 02, 03, 06, 07, 09, 11, 12, 13 |
| 2 (●) | Windings (146), cable (106), MRI & NMR (70), REBCO (69), smes (50), Nb$_3$Sn (49), tokamak (42), cryogenic (41), electric properties (35), motor (35) | 38 | 20.08 ± 28.59 | 01, 03, 04, 06, 07, 09, 11, 12, 13, 14 |
| 3 (●) | DFT (67), electronic structure (40), optical (38), spintronic (38), Heusler alloy (27), first-principles cal- culation (25), ferromagnetic (23), anisotropy (21), electronic property (19), half-metallic (18) | 23 | 20.39 ± 13.75 | 03, 06, 07, 09, 11, 12, 13 |
| 4 (●) | Nano (95), biomedical application (28), graphene (19), nanowire (19), ferrite (18), polymer (15), superparamagnetic (14), co-precipitation (8), electromagnetic (8), Nanostructure (8) | 10 | 23.20 ± 24.66 | 02, 03, 06, 07, 09, 11, 12, 13 |
| 5 (●) | XRD (44), sol-gel (18), crystal structure (14), Raman (13), dielectric property (12), spinel ferrite (10), dielectric (7) | 7 | 16.85±11.52 | 03, 06, 07, 09, 11, 12, 13 |
| 6 (●) | Accelerator (67), normal-conducting (12), magnetic devices (11), magnetic permagnets (10), radiation hardened magnet (10), wiglers and ondulators (10) | 6 | 20.00 ± 21.03 | 07, 09 |
| 7 (●) | Perovskite (30), ferroelectric (12), oxide (10) | 3 | 17.33± 8.99 | 03, 06, 07, 09, 11 |
| 8(●) | Material (15), metamaterial (14), terahertz (14) | 3 | 14.33 ± 0.47 | 03, 07, 09 |
| 9 (●) | fault current limiter (33), hvdc (9) | 2 | 21 ± 12 | 07, 09, 11 |
| 10 (●) | Electron-phonon (12), Hydrides (10) | 2 | 11.00±1.00 | 01, 03, 07, 09, 13 |
| 11 (●) | Thermal (10), Transport (10) | 2 | 10.00±0.00 | 03, 07, 08, 09, 12, 13 |
| 12 (●) | magneticocaloric effect (10), curie temperature (7) | 2 | 8.5±1.5 | 06, 09, 13 |
| 13 (●) | Conductivity (15) | 1 | - | 03, 06, 07, 09, 11 |

$^a$ Abbreviation: number of keywords (NK), average frequence (AF), standard deviation of AF (SDAF)

these advancements in aviation technology contribute to the global efforts to combat climate change and promote sustainable industrialization. Finally, using superconducting materials for massive transporta- tion applications reflects the potential to achieve SDG 07 (Affordable and Clean Energy) by developing innovative and sustainable energy solutions.

Lastly, it's important to underscore the crucial role of Superconductivity in promoting SDG 7 (i.e., Affordable and Clean Energy). By improving power transmission, generation, and storage, Superconduc- tivity leads to greater energy efficiency and significantly reduces environmental impact. This is a beacon of hope in our global efforts towards sustainable development. Furthermore, superconductors are actively contributing to the development of Industry, Innovation, and Infrastructure (i.e., SDG 9) by enhancing the efficiency and sustainability of various technologies. Lastly, the development of new superconducting materials applied to the energy field also supports the pursuit of SDG 12 (a transversal SDG, Responsible Consumption and Production) by reducing energy waste and improving resource utilization.

To date, the most incredible human and environmental challenges are still unsolved. Hence, bold actions are needed to overcome them, which is where the SDGs come in. In addition, specific keywords emerged in the 2000s and showed an increasing trend through the SDGs stage (see Figure 12). Although this work focuses on superconducting applications directly linked to the SDGs, it is important to note that basic research is extremely important for expanding human knowledge and improving our ability to address and solve problems. Some of these solutions can lead to new technologies and applications, enabling us to fulfill the SDGs. In the field of superconductivity, efforts have been made to understand the fundamental mechanisms of unconventional superconductors [112, 113], leading to the discovery of new materials such as nickelates [114–119]. Additionally, with further development, topological supercon- ductivity [120–123] could be applied in supercomputers. BCS materials under Earth's core pressures have also achieved near-room-temperature critical temperatures [124], something never accomplished before.

Returning to the focus of the present work, the following section explores prominent trend topics in superconductivity, superconductors, and their applications within the contexts of health, energy, quantum sensors, and sustainable transportation, all framed by the SDGs.

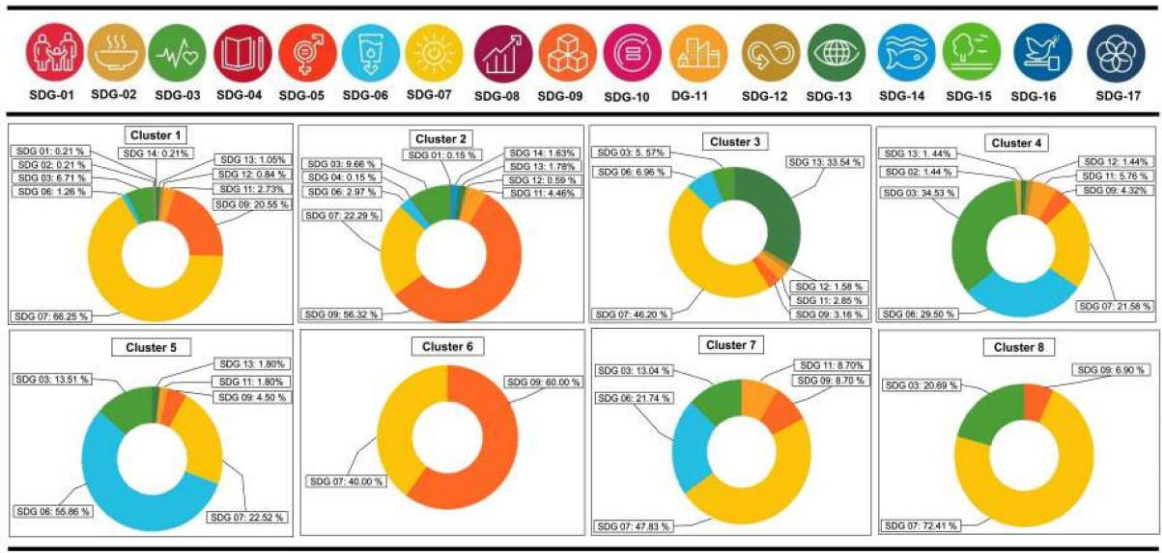

Figure 11: Keywords distribution per SDG from 2015 to 2023.

# 6 Superconductivity: Properties and Classification

The origins of low-temperature physics are rightly attributed to the liquefaction of helium in 1908 and the serendipitous discovery of superconductivity in mercury. These pioneering achievements are credited to the Dutch physicist Heike Kamerlingh Onnes in 1908 and 1911, respectively [125]. Superconductivity is a quantum phenomenon characterized by zero electrical resistance and the expulsion of magnetic fields in certain materials when cooled below a critical temperature $T_c$ [126]. Materials exhibiting these properties are known as superconductors (SC) [127, 128]. Therefore, any superconducting material must demonstrate three key properties: (i) zero resistivity, (ii) perfect diamagnetism (the Meissner effect), and
(iii) a discontinuity in specific heat at $T_c$, showing an exponential behavior well below such a temperature [129–132].

The first property, zero resistivity, was initially observed by Onnes, who noted that mercury, when cooled with liquid helium, exhibited a drastic decrease in resistivity, reaching zero at 4.15 K, indicating the transition to superconductivity. Consequently, superconductors display zero resistance to direct electric currents up to a critical value, beyond which they return to a normal state. This property allows superconducting materials to eliminate power loss due to the Joule effect [129, 133, 134].

The second property involves the Meissner effect, discovered by W. Meissner and R. Ochsenfeld in 1933 [130, 135, 136], which results in the complete expulsion of magnetic flux from the interior of super- conductors when cooled below $T_c$ in the presence of a static magnetic field (field cooling, FC procedure). The third property, first described by Landau in 1937, pertains to the specific heat of superconductors and forms the basis of the Ginzburg-Landau Theory (GLT) [137]. In superconducting materials without an applied magnetic field, the transition from normal to superconducting state does not involve latent heat, marking it as a second-order phase transition. This behavior is characterized by a discontinuity in the specific heat curve at $T_c$ [138–141].

Today, a wide range of materials exhibiting superconducting states, can be classified by criteria such as $T_c$ [138], response to magnetic fields [136, 142, 143], material composition [144–147], and the theoretical frameworks governing their behavior [131, 137, 148]. Based on their critical temperatures, superconductors are categorized into two main types: high-temperature superconductors (HTS), which can operate above 77 K and include materials like yttrium barium copper oxides (YBCO) [140] and bismuth strontium calcium copper oxides (BSCCO) [149], and low-temperature superconductors (LTS), such as niobium-titanium (NbTi) and niobium-tin ($Nb_2Sn$) alloys, which must operate at temperatures below 15 K and thus, require liquid helium cooling [150]. LTS are commonly employed in coils to generate high magnetic fields (tens of Teslas). Conversely, applications using HTS still encounter significant challenges.

Typically, type-I superconductors are LTS, exhibiting an abrupt transition to the superconducting state and excluding all magnetic fields from their interior. Type-II superconductors, encompassing both LTS and HTS, exhibit a more gradual transition and allow limited (quantized) magnetic field penetration, forming magnetic vortices [138]. Due to their higher critical magnetic fields and possible current densities, type-II superconductors are favored for current and future applications. The specific properties of these superconductors dictate their potential applications, a topic to be explored further in the subsequent section of this review.

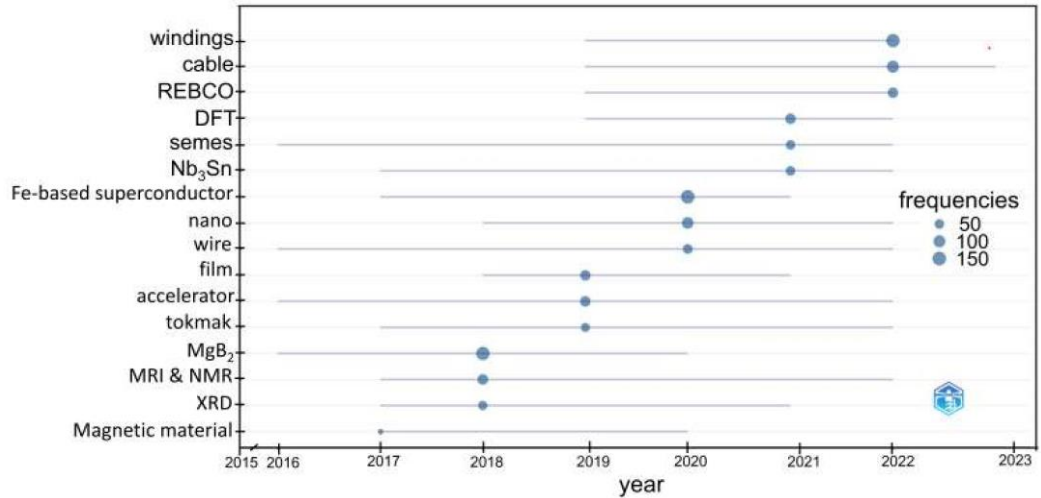

Figure 12: Trend Topics on the superconductivity, superconductors and their applications during the SDG stage (source: bibliometrix software).

# 7   Applications of superconductivity to achieve the SDGs

It took considerable time after the discovery of superconductors to identify materials suitable for practical applications. In 1986, Bednorz and Müller [150] discovered HTS superconductivity in the La- Ba-Cu-O system, which forms a class of copper oxides. The HTS are often referred to as non-conventional superconductors, as an explanation using the BCS theory is not possible and a complete explanation of the superconductivity is still lacking. In Figure 1 of Ref. [151] it is presents $T_{c,max}$ of several families of copper- based HTS (La-based, Y-based, Bi-based, Tl-based and Hg-based) in ambient conditions. The highest superconducting transition temperature found so far in the copper oxides is up to 138 K (164 K under pressure) in the Hg-family [151]. These HTS materials sparked an intense effort to develop applications that could operate at much higher temperatures, such as the cheaper and more easily attainable boiling point of liquid nitrogen (77 K) [152]. In 2006, another different class of HTS superconductors was found, based on Fe instead of Cu, also called iron-based superconductors (abbreviated IBS or FeSC). Figure  1 of Ref. [151] includes the temperatures of the important coolants, LHe (4.2 K), $LH_2$ (20.7 K) and $LN_2$ (77 K) as violet dashed lines. The materials with the highest $T_c$ could even operate in the night on moon without an additional coolant. Furthermore, it can be seen that some HTS materials require LHe as coolant, but to distinct them from LTS, these materials show large critical fields. Cooling with $LH_2$ is especially useful for the IBS superconductors, which do not reach the 77 K line. However, the emergence of HTS materials underscored that $T_c$ alone is not the sole crucial parameter for practical superconductors. Thus, other factor like strong anisotropy of $j_c$, granularity and grain boundaries, low irreversibility lines and also toxicity of the raw materials (e.g., Tl, Hg or As) may play an important role for decision to use HTS materials for applications.

The focus of this section is on the applications of superconductors to achieve the Sustainable Develop- ment Goals (SDGs). Nevertheless, superconductors face traditional constraints and challenges inherent in the design, synthesis, and manufacturing of new materials [53, 153–156]. Practical superconductors must reliably carry large currents, typically characterized by a critical current $I_c$ or critical current density $J_c$, in significant magnetic fields, which also poses an upper operating limit known as the upper critical magnetic field $H_{c2}$. These parameters are interrelated; each is a function of the others below their critical values. This is illustrated in Fig. 13, showing the 3D surface of $j_c(T, B)$ for a type-II superconductor. In practical applications, $H_{c2}$ often does not independently limit the use of superconductors because, in power applications for example, the magnetic field is typically generated by the current flowing in the superconductor itself. Thus, the design limitations stem from $T_c$ or, more commonly, $I_c$, which is influenced by the magnetic field $H$ at the superconducting wire and vice-versa. Engineering constraints in working with superconductors are particularly significant. Safety margins are crucial for machine reliability and longevity, necessitating that designers typically operate machines at a fraction (e.g., half) of $I_c$ and $T_c$, rather than their theoretical maximum capability or "entitlement" [152, 157].

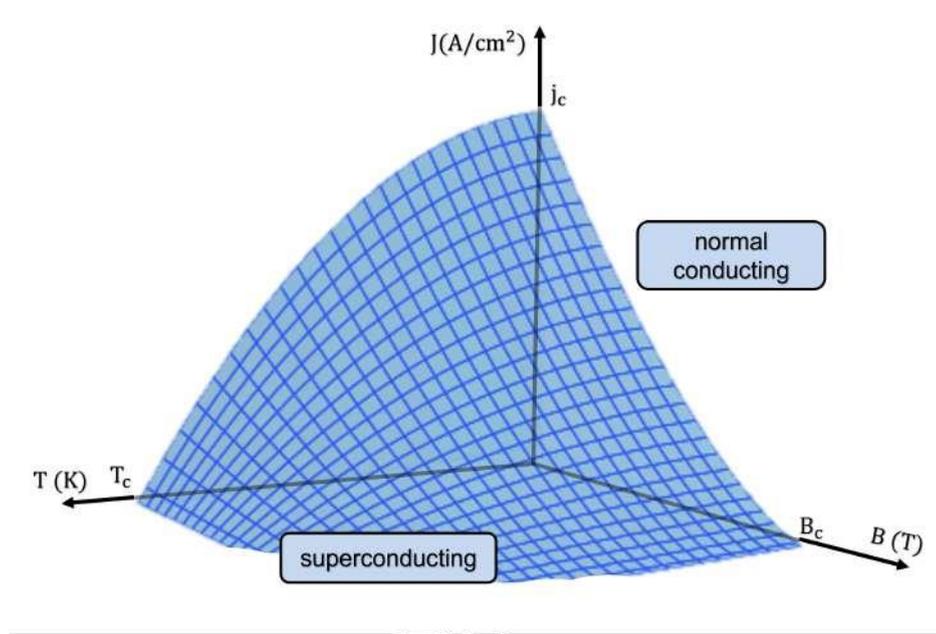

Figure 13: The 3D surface of $I_c(T, B)$ in type-II superconductors.

Now, it is important to mention the superconducting materials which are considered for applications. Although in the meantime a myriad of superconducting compounds were discovered, the current existing and planned applications of superconductivity focus mainly on a small number of materials which have proven to be the best and sturdiest materials. These superconducting materials are the LTS materials NbTi, Nb₃Sn for superconducting wires and coils, Nb and NbN for quantum applications and sensor elements. All these materials require the cooling by liquid Helium.

Speaking about higher operation temperatures, which enable a more cost-effective cooling by liquid hydrogen (20.3 K) or liquid nitrogen (77 K), only 5 HTS materials remain, representing the most re- searched superconducting materials for possible applications. The crystal structures of these materials are presented in Fig. 14, including the *RE*BCO (or often called *RE*-123) system, were *RE* stands for rare earths like Nd, Eu or Gd. The ionic radius of these rare earths is much larger than that of the original Y, leading to the interesting effects enhancing both the achievable critical currents and critical fields [158, 159]. The superconducting transition temperature of these systems range between 91 and 95 K, so well above the boiling point of liquid nitrogen. Thus, the 123-type superconductors are the current material of choice for high-current applications in form of so-called coated-conductor tapes, where the alignment of the superconducting grains became possible. The two Bi-based materials, Bi₂Sr₂CaCu₂O₈ (or Bi- 2212) and Bi₂Sr₂Ca₂Cu₃O₁₀ (or Bi-2223) are members of one large family, having one (Bi-2212) and two (Bi-2223) Cu-O-planes. The superconducting transition temperatures are 85 K and 105 K, respectively. Both materials were employed for the first generation of superconducting tapes due to their very high upper critical fields at low temperatures and their easier crystal alignment, allowing the production of powder-in-tube tapes and wires. The SmFeAsO material (abbreviated as 1111) is the parent compound of the iron-based superconductors (IBS), found in 2006. The material being best suited for applications is the IBS 122-family, represented here by Sr/BaFe₂As₂ with a $T_c$ of 55 K. From this compound also wire and tapes can be produced effectively. Although the superconducting transition temperatures of the 122-family are lower than 55 K, the cooling with cryocoolers to about 20 K and especially, with LH₂ is an important issue for applications as the critical fields are high and no expensive rare earth materials are involved. FeAs (abbreviated 11) presents the nominally simplest HTS material with only two elements being involved. The $T_c$ of the bulk material is only 8 K, but can be increased by replacing half the Se by Te to 15 K, and in form of monolayers, $T_c$ can be above 65 K [160].

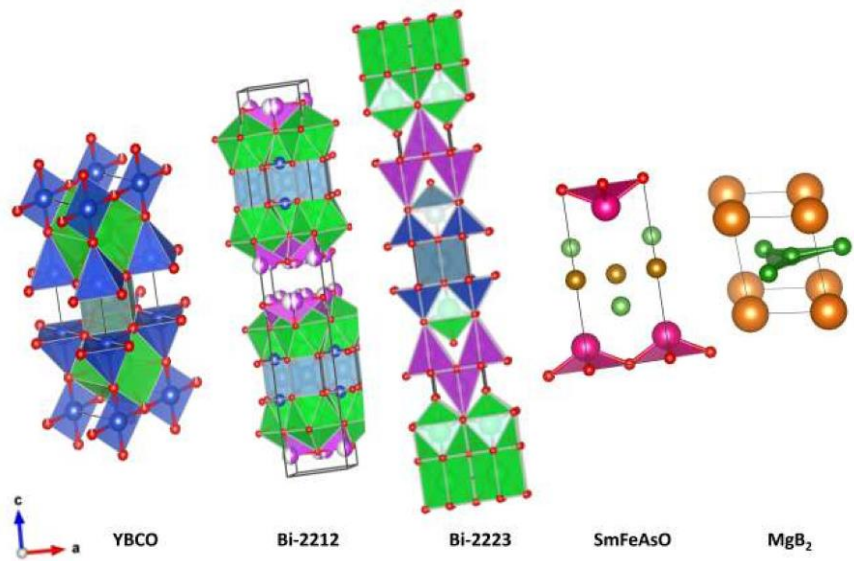

Figure 14: Crystal structures of the most important HTS materials for applications and the metallic superconductor, $MgB_2$. All structures werre drawn using the software VESTA [161]

The last material shown in Fig. 14 is $MgB_2$, the metallic superconductor with the highest $T_c$ known: 39 K. This material was a surprise discovery in 2001 [162], leading to another research peak after the discovery of the HTS. All the research efforts now enable the production of long wires of $MgB_2$ in a manner analog to the fabrication processes applied for NbTi and $Nb_3Sn$. Like in the case of the latter two materials, the grain orientation is not playing a major role for this superconductor, which enables a well-known and cheap production technology to be employed [163, 164]. Furthermore, $MgB_2$ does not contain rare earth elements, is relatively cheap and a lightweight material, which may also be an important issue for some applications.

It is important to note that all the 5 materials shown in Fig. 14 exhibit characteristic planar structures (either Cu-O or Fe-O, and hexagonal B in the case of $MgB_2$). These planes are the essential highway of superconductivity. Only the Y/*RE*-family also includes Cu-O-chains as a second element. Thus, as noted in Bray's review [152], the HTS materials exhibit high anisotropy (planar orientation) and have short coherence lengths. These characteristics necessitate that the materials have their major $(a-b)$ planes well- aligned (textured) to achieve low-angle grain boundaries, which are crucial for efficient current transport between the superconducting grains. Achieving this alignment in long superconductor wires ranks as the second most challenging task after producing kilometer-long single crystals, and it has been a focal point of research and development efforts in superconductor wire production since 1986, continuing to present times.

The Figure 1 if Ref. [165], summarizes the upper critical fields, $H_{c2}(T)$ (full lines) and the irreversibil- ity fields, $H_{irr}(T)$ (dashed lines) of the practical superconductors for applications. The irreversibility field defines where the flux pinning goes to zero, which sets the limit for any practical application of a super- conducting material. Remarkably is the difference between $H_{c2}(T)$ and $H_{irr}(T)$ for the YBCO system, which reflects the effects of thermal fluctuations, being a large problem for applications at elevated tem- peratures. Here, one can see that low temperatures, the conventional materials NbTi and $Nb_3Sn$ are beaten by $MgB_2$ as well as $FeSe_xTe_{1-x}$ (the 11-IBS-compound) with higher $H_{c2}$-values, whereas all other HTS materials show even larger $H_{c2}$-values, which may be > 100 T. One must note here that both BSCCO compounds show quite low $H_{c2}(T)$-lines, which strongly increase at temperatures below 20 K. This directly implies that the first generation tapes using the Bi-2223 compound can hardly be used for coils operating at 77 K, but work well in low-field situations (e.g., cables). This low $H_{c2}(T)$ at 77 K is a direct consequence of the 2D-character of these materials having a layered, strongly anisotropic crystal structure. In contrast to this behavior, $H_{c2}(T)$ of the *RE*-123 compounds works well at 77 K and it is thus the material of choice for the second generation of superconducting tapes, the so-called coated conductors. Cooling with $LH_2$ brings $MgB_2$ and the IBS 1111 and 122 materials into consideration, which are well suited to work at this temperature as indicated by their $H_{c2}(T)$.

Moreover, integrating sustainability considerations adds another layer of complexity to the selection of specific superconducting materials for practical applications. Therefore, in the following section, we will discuss key aspects related to the sustainability of modern superconductor applications.

## 7.1    Applications in the field of human health

The advancement of human development has driven various research fronts and technological fields globally throughout history. When focusing on healthcare, numerous applications leverage the phe- nomenon of superconductivity. In medical contexts, superconductors' magnetic properties play a pivotal role, particularly in magnetic resonance imaging (MRI) for diagnostic purposes, utilizing nuclear mag- netic resonance (NMR) and electron paramagnetic resonance (EPR) techniques [166, 167]. Additionally, superconductors are employed in radiotherapy and in the production of isotopes for medical applications [168].

Superconductivity, initially a scientific curiosity, has transformed into a catalyst for improving quality of life, notably through the invention of MRI in the late 1970s. This phenomenon significantly enhances the feasibility of MRI, leading to the installation of over 40,000 superconducting MRI scanners in hospitals worldwide [169].

Nuclear magnetic resonance (NMR) imaging is a non-invasive technique based on the principles of atomic behavior and physical properties. It involves a powerful magnetic field and radio frequency pulses to generate detailed images of soft tissues such as muscles, organs, joints, and the nervous system [170]. During the procedure, the patient lies within an MRI scanner, exposed to a strong magnetic field generated by a superconducting magnet capable of operating at 1.5 and 3 T [171, 172], or even higher fields [172]. In this environment, atomic nuclei, particularly hydrogen protons in the body, align themselves along the magnetic field's direction. The imaging process captures changes in the alignment of hydrogen atoms as they emit absorbed energy upon returning to their original state [173]. These signals are then processed by computers to construct detailed images of internal tissues [171].

The quality of Magnetic Resonance images is influenced significantly by several factors, with the uniformity of the magnetic field being of utmost importance [174]. A homogeneous magnetic field is critical because any lack of uniformity can introduce image noise and imperfections [175, 176]. Therefore, the magnets used in MRI devices must achieve a high degree of uniformity, typically on the order of 10 parts per million peak to peak, with an approximate diameter of 50 cm [175, 176]. However, achieving greater uniformity while reducing the size of magnets poses challenges as it can lead to larger and heavier MRI scanners [177].

Recent advancements in manufacturing superconducting magnets have focused on increasing field strength, but this has been accompanied by greater energy demands and larger sizes over the past few decades, as discussed by Jimeno *et al.* [170]. Conversely, efforts in optimizing algorithms and designing superconductor wires have progressively reduced magnet length and weight, resulting in lighter and more compact scanners [178]. Jimeno's review also highlights global disparities in MRI technology distribution, with variations in MRI density across different regions. Low- and middle-income countries often use low- intensity field systems based on permanent magnets, which are associated with lower acquisition and maintenance costs.

Therefore, there is a concerted effort towards developing new NMR magnets with reduced energy con- sumption and production costs, aligning with the SDGs aimed at improving global healthcare accessibility and sustainability.

Over the years, companies like Varian/Oxford and Bruker have developed various NMR spectrometers equipped with superconducting magnets, continually increasing their operating frequencies. Bruker, for instance, explored Bi-2212 and Bi-2223, and ultimately settled on *REBCO* conductors around 2010 to achieve proton resonance frequencies well beyond 1.0 GHz [179]. Additionally, Bruker pursued a 1.2 GHz magnet using a hybrid HTS/LTS design to minimize HTS use and technological risks.

The US National High Magnetic Field Laboratory (NHMFL) also made significant advancements by developing a series-connected resistive/superconducting hybrid (SCH) magnet capable of reaching fields up to 1.5 GHz. This magnet was specifically designed for enhanced stability and homogeneity in the tenths of ppm range crucial for high-resolution NMR spectroscopy. Coupled with a Bruker AVANCE NEO console, it opens new frontiers in NMR research at these field strengths prior to widespread availability of HTS magnets [180]. Recently, there have been reports on preliminary designs for a 1.8 GHz NMR magnet based on LTS/REBCO systems [181].

Traditionally, NMR employs superconducting magnets based on Niobium Titanium (NbTi), requiring cooling with liquid helium to achieve superconducting states. While this setup is economically efficient and ensures optimal magnetic field homogeneity, it presents challenges as the magnet and cryogenic system account for approximately 38% of the total cost [182]. Recent research efforts have focused on modifying cooling systems to reduce costs and explore HTS like magnesium diboride ($MgB_2$) [183, 184], yttrium barium copper oxide (YBCO) [185, 186], rare earth barium copper oxide (REBCO) [187, 188], and strontium bismuth calcium copper oxide (BSCCO) [189, 190]. These materials, due to their higher critical temperatures, eliminate the need for liquid helium, thereby lowering operating costs. This aspect holds promise for enhancing healthcare accessibility, particularly in underserved regions of developing countries.

In the field of radiotherapy, the use of energetically charged particles, such as heavy ions, offers an effective means to precisely target and deliver doses to deeply located tumors in patients. The energy distribution of these particles helps minimize radiation exposure to surrounding healthy tissues, thereby enabling patients to complete oncological treatments with fewer side effects. Heavier ions, like carbon, exhibit reduced lateral scattering and enhanced biological effectiveness compared to lighter particles [191, 192].

Despite these advantages, the use of heavy ions in radiotherapy is constrained by the size and com- plexity of the synchrotron accelerators required to generate them. Addressing these challenges involves ongoing research into novel superconducting ceramics aimed at shrinking the size and improving the efficiency of synchrotron technology [191–194].

Recent advancements in radiotherapy research have introduced a promising technique known as Fast Low Angle Shot (FLASH). This approach shows potential in minimizing damage to normal tissues while maintaining or even enhancing anti-tumor efficacy compared to conventional radiotherapy methods. A significant development in this area is the establishment of the Advanced Radiotherapy Research Plat- form (PARTER) in 2019. The PARTER platform employs megavoltage X-rays for experimental studies using the FLASH technique. Central to the operation of PARTER is the utilization of superconducting materials [195].

Superconductors offer distinct advantages due to their ability to operate without significant energy losses and maintain electrical stability under extreme conditions. In the context of PARTER, the super- conducting linac plays a pivotal role in achieving ultra-fast dose rates necessary for FLASH radiation. Superconductivity enables higher currents and faster dose delivery, which are critical for realizing the specific benefits of FLASH therapy— preserving healthy tissues while effectively treating tumors [196].

In the realm of hadron radiotherapy, heavy ions accelerated by superconducting or conventional magnets are utilized to produce a highly localized dose delivery within a patient's body, known as the Bragg peak. This precise targeting enables the concentration of radiation within tumors while minimizing exposure to surrounding healthy tissues. The first facility for heavy ion beam radiation therapy was established at the Lawrence Berkeley National Laboratory in 1957. Since the 1990s, hadron therapy has evolved into a widely practiced medical treatment, with numerous facilities worldwide adopting this advanced technique. Superconducting magnets have increasingly played a crucial role in accelerating particles and guiding their beams to achieve optimal delivery precision [169].

In pharmaceutical applications, the precise delivery of pharmacologically active substances is paramount to avoid unintended side effects and optimize therapeutic outcomes. The development of innovative Drug Delivery techniques is pivotal in this regard. Superconductor composites have emerged as promising materials due to their ability to generate intense magnetic fields, facilitating the magnetic transport of drugs to targeted locations within the body.

Superconducting drug delivery offers a targeted approach that minimizes systemic side effects and maximizes treatment effectiveness. Furthermore, integrating superconductor technology into controlled release systems allows for the modulation of drug release in response to external magnetic stimuli, thereby tailoring treatment to individual patient needs. This capability represents a significant advancement in pharmaceutical science, promising improved patient outcomes through enhanced precision and efficacy in drug delivery [195, 197].

Superconductivity has revolutionized medical diagnostics and healthcare through various innovative applications, significantly impacting the achievement of the SDGs. One of the key applications is Mag- netoencephalography (MEG), which utilizes Superconducting Quantum Interference Devices (SQUIDs) to measure the magnetic fields generated by brain activity with exceptional sensitivity [198, 199]. This technique provides crucial insights into neural activities, aiding in the study of neurological disorders and functional brain mapping.

Similarly, Magnetocardiography (MCG) employs SQUIDs to detect magnetic fields associated with cardiac activity, offering a non-invasive and highly sensitive approach to diagnosing cardiac conditions [200–202]. This capability is particularly valuable for diagnosing arrhythmias and other cardiac disorders early in fetuses, thereby contributing to better maternal and fetal health outcomes.

Furthermore, SQUIDs have been instrumental in biomagnetic applications beyond MEG and MCG,

enabling precise measurements of magnetic fields from specific organs and physiological processes [203, 204]. This precision enhances our understanding of human physiology and facilitates the development of advanced diagnostic methods.

In addition to SQUIDs, Superconducting Nanowire Single Photon Detectors (SNSPDs) have emerged as a promising technology in medical diagnostics. These detectors, capable of detecting single photons with high sensitivity, are used in various biological and physiological systems [205–207]. They are par- ticularly effective in applications such as oxygen luminescence detection and Fluorescence Correlation Spectroscopy (FCS), where sensitivity and temporal resolution are critical [208]. Moreover, SNSPDs demonstrate exceptional performance in the infrared and soft X-ray ranges, making them versatile tools for advanced medical imaging applications.

The impact of superconductivity in medical diagnosis aligns closely with SDG 3 (Good Health and Well-being). By enhancing diagnostic accuracy, reducing the invasiveness of procedures, and improving patient outcomes, these technologies contribute to promoting health and well-being globally. Moreover, the accessibility and reliability of these advanced diagnostic tools can help reduce health inequalities (SDG 10), ensuring that all individuals have access to equitable healthcare services.

In conclusion, superconductivity represents a transformative force in healthcare, driving innovations that support global efforts towards healthier and more sustainable societies. By enabling precise and non-invasive medical diagnostics, superconducting technologies contribute significantly to achieving SDGs related to health, equality, and sustainable development.

## 7.2    Applications in the Traffic

Magnetic levitation (Maglev) trains enable the achievement of high-speed and low-friction travel. Maglev trains are propelled by a system where magnetic coils along the tracks generate a powerful magnetic field. This field interacts with magnets on the train, causing it to levitate above the track without physical contact, thereby eliminating friction. The primary advantage of Maglev technology lies in its ability to achieve speeds well above conventional trains, often exceeding 500 km/h, while maintaining stability, reducing noise, and minimizing vibrations [209–211]. This efficiency is coupled with a significantly lower carbon footprint compared to traditional locomotives, making Maglev trains an attractive option for sustainable transportation solutions.

Despite these advancements, challenges remain, particularly concerning the stability of Maglev trains, especially when encountering track irregularities. Research, such as that by Zhai *et al.* [209], has high- lighted issues where irregularities can affect the suspension distance and potentially compromise ride quality at higher speeds. Solutions proposed include integrating accelerometers into the electromagnets to enhance train performance and stability.

All current existing Maglev operations use conventional coil systems for both levitation and propulsion. Superconducting materials have thus the potential to revolutionize transportation, particularly in the context of superconducting Maglev trains. In this case, the onboard magnetic system is replaced by superconducting magnets, as will be discussed in the next section. Overall, Maglev trains represent a significant application of superconductivity in transportation, offering a glimpse into a future where high-speed, energy-efficient, and environmentally friendly travel could become more commonplace, con- tributing to sustainable development related to infrastructure, innovation, and climate action. Maglev systems have been implemented and researched in countries like Japan, Germany, France, China, Brazil and Russia, reflecting their potential as a viable future transportation solution. Continued research and development in this area are crucial for further improving the reliability and operational efficiency of Maglev systems worldwide.

It must be mentioned here that the research on Maglev does not only consider high-speed transportation, but also includes systems like people movers (e.g., Supratrans), contactless working and moving in laboratories (FESTO) or even elevator systems in skyscrapers (Thyssen-Krupp Elevator).

Superconductors are paving the way for significant advancements in transportation, not only in high- speed trains but also in the aviation sector, particularly with efforts aimed at enhancing aircraft propulsion systems.

### 7.2.1    Maglev Trains and Superconductors

Figure 15 presents photos from the viewing center at the Yamanashi testline in Japan. The main photo was taken in July, 1998 with a passing MLX-01 train set. The inset shows the further development with improved aerodynamics (MLX01-901).

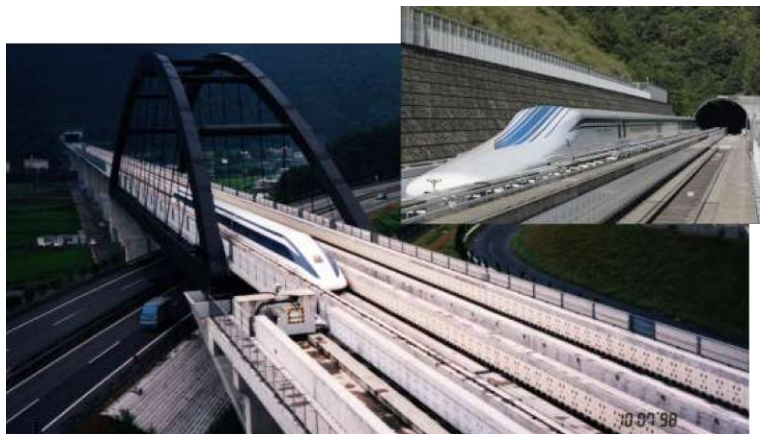

Figure 15: The Japanese MLX01 Maglev train at the Yamanashi test line. Inset: Maglev car MLX01-901 with improved aerodynamics.

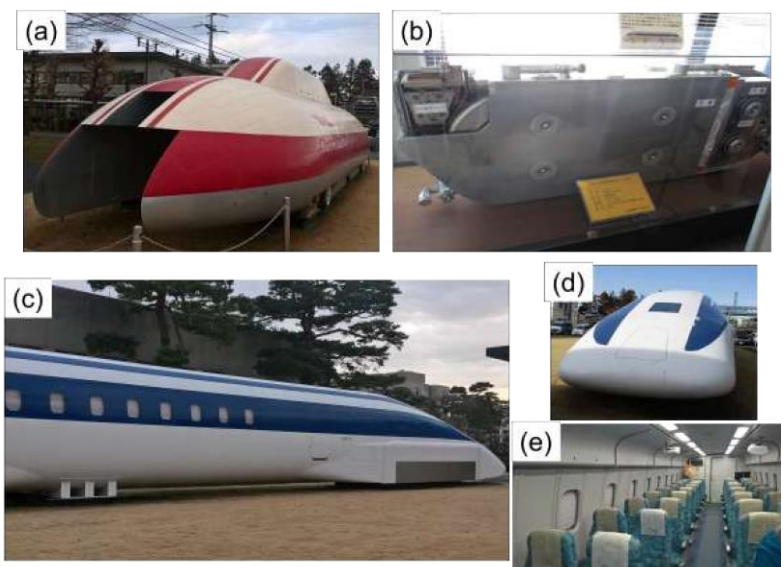

Figure 16: Maglev cars exhibited at RTRI in Kokobunji, Tokyo. (a) The very first superconducting levitation car, ML-500 with T-shape magnetic rail. (b) The cryostat with the superconducting coil system (NbTi) of the MLU002. (c) Side view of the car MLX01-3. The grey section in front is the location of a cryostat system like the one shown in (b), which is not present in the car. (d) Front view of MLX01-3, and (e) shows the interior of MLX01-3, which is now serving as an information center.

The first superconducting Maglev trains were developed in Japan, with a testline at Miyazaki pre- fecture starting from 1977. From 1996, a new testline at Yamanashi prefecture (west of Tokyo, close to Mount Fuji) started operation. This line is intended to be a part of the new Chuo-Shinkansen line between Tokyo and Osaka. All these first Maglev cars were equipped with superconducting coils based either on NbTi or $Nb_3Sn$, requiring onboard LHe-cooling. Several of these Maglev cars are now exhibited at the RTRI institute in Kokobunji, Tokyo, as shown in Fig. 16 (a–e).

Maglev trains using superconductors, such as those based on HTS like YBCO, have demonstrated compelling advantages, especially in terms of acceleration and operational stability at high speeds [210, 212]. Compared to traditional high-speed trains based on intense magnets, HTS-based maglevs offer superior performance characteristics that align well with the requirements of modern high-speed railway systems. These systems reflect their potential as a viable future transportation solution.

### 7.2.2 Superconductors in Aircraft Propulsion

In the aviation sector, there's a pressing need for more efficient propulsion systems to reduce pollution, noise, and fuel consumption. NASA's N3-X project exemplifies efforts to revolutionize aircraft design through the integration of superconducting technologies [213]. The N3-X concept, derived from the Boeing

B777-200LR, employs superconducting electrical generators and a distributed turboelectric propulsion system. This innovative configuration aims to achieve a substantial 70% reduction in fuel consumption by using superconductor-based synchronous motors with superconducting components in both rotor and stator assemblies [214–216].

From the perspective of materials for such applications: (i) $MgB_2$, despite its relatively low $T_c$ of 32 K, was chosen for its superconducting properties suitable for aircraft applications. Ongoing research focuses on improving its performance and minimizing losses. (ii) BSCCO (Bi-Sr-Ca-Cu-O) is actively being researched with a focus on reducing alternating current losses, making it viable for use in aircraft propulsion systems. On the other hand, (iii) the application of YBCO in aircraft propulsion remains under exploration, with efforts needed to effectively mitigate AC losses. As future directions, research initiatives are exploring the development of superconducting motors and hydrogen fuel cells for aircraft electric propulsion systems. These efforts aim to achieve higher efficiency and power densities, crucial for advancing sustainable aviation technologies [217].

In summary, superconductors are poised to revolutionize both land-based and airborne transportation sectors, offering solutions that enhance efficiency, reduce environmental impact, and propel technological innovation toward a more sustainable future. Continued research and development in superconducting materials and applications are pivotal for realizing these transformative advancements in transportation technologies.

Absolutely, the development and implementation of more efficient and environmentally friendly air- craft and Maglev technologies using superconducting materials align strongly with several Sustainable Development Goals (SDGs). Regarding SDG 13: Climate Action, advancements in aviation technology aimed at reducing fuel consumption and emissions by integrating superconducting technologies in aircraft propulsion systems, such as those explored in NASA's N3-X project [213], achieve substantial reductions in greenhouse gas emissions. These innovations help mitigate the aviation sector's environmental impact, contributing to global efforts to combat climate change. From SDG 09: Industry, Innovation, and In- frastructure, superconducting technologies in transportation, including aircraft propulsion systems and Maglevs, represent significant innovations in infrastructure and industry. These advancements enhance the efficiency and sustainability of massive transportation infrastructure. Moreover, such innovations contribute to building resilient infrastructure, promoting sustainable industrialization, and fostering in- novation essential for economic growth and development. Also, the application of superconducting mate- rials in transportation supports SDG 07 by promoting affordable and clean energy solutions. By reducing fuel consumption and improving energy efficiency, superconducting technologies contribute to sustainable energy use in the transportation sector [218].

In conclusion, the integration of superconducting materials in aviation technologies not only enhances operational efficiency and reduces environmental impact but also supports broader global sustainability goals outlined in SDGs 07, 09, and 13. These advancements represent pivotal steps toward achieving a more sustainable and resilient future for transportation infrastructure and industry worldwide.

## 7.3    Applications in the energy field.

One of the primary applications of superconductors contributing to the Agenda 2030 is in Affordable and Clean Energy (SDG 07). This application addresses issues related to the electrical transmission from power plants to urban centers and the cooling costs in small transmission stations near areas of electricity generation and consumption. However, due to the high cost and complexity of cooling superconducting cables over long distances, these applications are not economically viable for long-range transmission networks. Instead, they are recommended and used primarily in small transmission stations or regions close to generation facilities. For instance, cables up to 120 meters in length, capable of carrying 100 million watts, are currently available. These cables use the superconducting ceramic YBCO [219]. Given the current significance of superconducting cables for efficient electrical transmission in various power plants, this section will highlight the latest research in this field [61, 220, 221]. Superconducting transmission cables (SCTCs) are typically characterized by a transverse dimension of a few tens of centimeters, with a characteristic length of the cables larger than the transverse dimension by several orders of magnitude [222]. An interesting characteristic that allows HTS cables to be classified refers to the location of the dielectric, which can be inside the cryostat (cold dielectric, CD) or outside the cryostat (warm dielectric, WD, as well as single-core or multi-core cables for three-phase designs [222, 223].

Furthermore, this technology can provide excellent network stabilization and significantly reduce line losses in scenarios where shorter distance cables are imperative, such as in railway traffic [215]. These advancements highlight the promising potential for practical applications of superconducting cables across various sectors, contributing to more efficient and sustainable electrical energy transmission [61, 221].

Superconductors play a crucial role not only in the transmission and generation of energy but also in Energy Storage Technologies (EST). ESTs can enhance global and national electrical systems by providing stability, grid reliability, reduced physical space requirements, rapid response to demand, and minimal energy losses, among other benefits. Currently, two main types of such devices can be found: Superconducting Magnetic Energy Storage (SMES) and Flywheel.

An SMES system essentially consists of a device that stores direct current (DC) in a magnetic field [224]. Typically, this device includes a superconducting coil within a cryogenic system. Low-temperature superconductors (LTS) SMES are constructed from NbTi superconductors and use liquid helium coolant at a temperature of 4.3 K, while high-temperature superconductors (HTS) SMES are constructed from ceramic oxide superconductors and use liquid nitrogen coolant at 77 K [225]. Due to the absence of energy conversion from electrical to mechanical or chemical forms, the efficiency of an SMES system can be high [226]. Additionally, SMES offers advantages such as higher power density, faster charging and discharging capabilities, and a longer lifespan compared to other storage technologies [224, 227, 228].

Conversely to SMES, a flywheel device stores energy based on the Meissner effect. A flywheel is an inertia device that stores energy using the kinetic energy of a rotating mass, or rotor. The amount of energy stored depends on the mass of the rotor, its positioning, and the speed at which it spins [229]. In a flywheel, a cylindrical permanent magnet rotates at high speed over a superconductor, inducing a shielding current on it. This current creates a magnetic field that repels the magnet, allowing it to levitate and spin without friction [230–233]. The speed of Flywheel Energy Storage Systems (FESS) increases as energy is stored and decreases as energy is discharged. FESS can be classified into low-speed (spinning at less than 6,000 rpm) and high-speed (spinning at $10^4$-$10^5$ rpm). However, limitations of FESS include energy loss due to friction, and mechanical inefficiencies, which can reduce the system's overall efficiency [233].

One option to help overcome the limitations of SMES and FESS is the use of hybrid magnetic/superconducting energy storage systems. These systems aim to optimize the performance and efficiency of energy storage, ensuring a quick and stable response to energy fluctuations. Employing hybrid magnetic/superconducting energy storage systems increases energy storage capacity, improves energy stability and quality, reduces energy losses, and extends battery life [229].

In addition to enhancing energy storage, advanced superconducting technologies play a critical role in protecting electrical transmission and distribution systems. Short circuits caused by lightning and accidental contact between lines or the ground can induce fault currents, significantly increasing the electric current in a local network and damaging electrical equipment. To safeguard against these fault currents, special devices known as fault current limiters (FCL) are installed in the transmission network [234–237].

Typical FCLs, such as conversion line reactors, have high AC losses and can produce a voltage drop in the network during a fault current event. However, HTS technology offers a much better solution to the fault current problem and is one of the most successful applications of cuprate superconductors [238, 239].

In HTS-based FCLs (HTS-FCLs), a fundamental property of superconductors is utilized: the transi- tion from a zero-resistance superconducting state to a normal resistive state when the electrical current exceeds the material's critical current [240]. Schematically, the HTS-FCL consists of a coated conductive material with layers of HTS material (e.g., BSCCO ceramic) within layers of resistive materials. Under normal conditions, the current flows through the HTS layers in the FCL. When a fault occurs, the current exceeds the critical current of the HTS material, causing the HTS layers to transition to the normal state. The current is automatically diverted within milliseconds to flow through the higher resistance layers, effectively quenching the fault current amplitude. Fast-operating HTS-FCLs significantly reduce damage to electrical equipment caused by system failures [241, 242].

Finally, superconducting materials find application in various forms such as tapes, bulk materials, and cables [243, 244], particularly in power generators like wind turbines. Superconducting generators offer advantages such as increased efficiency, reduced maintenance costs, and smaller size compared to conventional generators [245–247]. These superconducting generators are also used in high-power density rotary motors, which are increasingly in demand for aircraft and ships in propulsion systems [217, 248– 250].

For industrial applications, coated superconductors formed from the HTS material *RE*BCO have a high transition temperature, and their current density in high magnetic field environments is higher than that of other superconductors. This high transition temperature offers the advantage of cooling the material without liquid helium, which could provide a stronger magnetic field at a lower cost. The applications of HTS magnets have attracted worldwide attention, and one such application is HTS linear motors.

These motors can be categorized into HTS linear synchronous motors (HTS-LSM) and HTS linear asynchronous motors (HTS-LIM). HTS-LSM motors, known for their high thrust density and efficiency, hold significant promise for industrial applications and engineering advancements [251].

One notable application of HTS-LSM motors is in high-speed and ultra-speed Maglev train traction motors. Recent research efforts have focused on the design, manufacturing, and testing of HTS magnets used as the motor magnet in HTS-LSM systems [251]. These magnets are designed as monopole HTS magnets, with coils manufactured using epoxy impregnation involving primary and secondary curing pro- cesses. Comprehensive static and dynamic tests have been conducted to thoroughly assess the magnet's characteristics [251].

This application highlights how superconducting technology, particularly in the form of HTS mate- rials, is advancing propulsion systems in transportation and industrial sectors, contributing to efficiency improvements and technological innovation.

These applications of superconductors in the energy sector strongly align with several Sustainable Development Goals (SDGs). Superconductivity plays a pivotal role in advancing SDG 7 (Affordable and Clean Energy) by enhancing the efficiency of power transmission, generation, and storage. This improvement contributes to greater energy efficiency overall and helps mitigate the environmental impact associated with traditional energy systems

Moreover, superconductors contribute significantly to SDG 9 (Industry, Innovation, and Infrastruc- ture) by fostering innovation and enhancing the sustainability of various technologies within the energy sector. This includes advancements in power generation, transmission systems, and energy storage solu- tions, which are critical for developing resilient and sustainable infrastructure

Additionally, the development of new superconducting materials applied in the energy field supports the objectives of SDG 12 (Responsible Consumption and Production). By reducing energy waste and improving resource utilization, these innovations promote more efficient and sustainable practices in energy production and consumption.

In essence, superconductors are integral to achieving these SDGs by driving technological advance- ments that optimize energy systems, enhance infrastructure resilience, and promote responsible resource management in the pursuit of a sustainable future.

## 7.4    Applications in sensors and quantum devices

Quantum Devices (QD) are instruments that exploit the principles of quantum mechanics to perform specialized functions. Among these, Quantum Sensors (QS) specifically detect and measure physical properties at the quantum level or monitor changes in the quantum environment, converting this data into signals or usable information. In recent years, superconducting devices have prominently emerged within QS, particularly in fields like quantum optics and quantum information processing.

Superconducting detectors offer unparalleled sensitivity in detecting infrared photons, facilitating significant advancements in quantum optics [252–254]. They also play a crucial role in developing secure quantum communication networks [255] and enable on-chip optical quantum information processing [256]. Furthermore, superconducting circuits based on Josephson junctions serve as a foundation for scalable quantum computing, extending into microwave wavelengths [253].

In the realm of quantum sensors, two types stand out: SQUIDs (Superconducting Quantum Interfer- ence Devices) and SNSPDs (Superconducting Nanowire Single Photon Detectors). SQUIDs, discovered over 50 years ago, remain vital for their ability to detect extremely weak magnetic field signals. Their applications span diverse fields such as biomedical analysis, where they serve as highly effective biosen- sors [204]. Moreover, SQUIDs have made significant contributions to geophysical research, aiding studies related to minerals and environmental resources .

A SQUID is constructed from one or more superconducting loops interrupted by Josephson junctions [257]. A Josephson junction consists of a thin layer sandwiched between two superconductors. When a voltage is applied across the junction, it allows a supercurrent to flow through it, influenced by an external magnetic field. Thus, when a magnetic field penetrates the superconducting loop of a SQUID, it induces a change in the supercurrent passing through the Josephson junction. This alteration results in a detectable change in the voltage across the junction, enabling measurement of the magnetic field strength. SQUIDs are exceptionally sensitive, capable of detecting magnetic fields as weak as a few femtoteslas (fT) [258], equivalent to those produced by a single neuron firing in the brain. This high sensitivity renders SQUIDs invaluable across biomagnetism, geophysics, and materials science.

Research into SQUIDs continues to improve their sensitivity and reliability. These advancements not only enhance their effectiveness in mineral analysis on Earth but also position them for potential applications in space exploration [203, 259, 260]. Ongoing developments in SQUID technology expand their potential across various fields, solidifying their role as versatile and indispensable tools. For instance, Oladapo et al. [257] describe advancements in SQUID sensor construction using pulsed laser deposition (PLD) of superconducting thin films on copper substrates in an ablation chamber, aiming to optimize sensor performance characteristics while reducing costs [257].

Halbertal *et al.* [261] present a nanothermometer based on a superconducting quantum interference device with a diameter of less than 50 nanometers that resides at the apex of a tapered pipette, which provides cryogenic thermal scanning detection that is four orders of magnitude more sensitive than previous devices, below 1 $\mu$K Hz$^{-1/2}$. This non-invasive and non-contact thermometry allows for obtaining thermal images of very low intensity and dissipating nanoscale energy up to the fundamental Landauer limit of 40 femtowatts for continuous reading of a single qubit at one gigaherz at 4.2 K. These advances enable observation of changes in distribution due to the charge of a single electron from individual quantum points in carbon nanotubes. It also reveals a dissipation mechanism attributable to resonant states located in graphene encapsulated within hexagonal boron nitride, opening the door to direct thermal imaging of nanoscale dissipation processes in quantum matter.

A SNSPD (Superconducting Nanowire Single Photon Detector) consists of superconducting nanowires with a flowing current close to their $J_c$. When a photon is absorbed by the nanowire, it disrupts the superconducting state, resulting in a detectable change in electrical resistance [262]. SNSPDs represent an advanced type of quantum device that surpasses other similar detectors, such as those based on semi- conductors. Their exceptional efficiency and performance are fundamentally attributed to key properties such as high detection speed and photon count rates reaching hundreds of MHz. These distinctive char- acteristics can be precisely tuned by selecting appropriate materials and optimizing the interaction with photons [263–265].

SNSPDs have catalyzed significant advances across various technological fields due to their unique properties. In quantum information science, these detectors play a pivotal role. Their high sensitivity and temporal resolution enable precise detection of individual quantum states, making them indispensable for applications such as quantum computing and secure quantum key distribution [266–268]. Moreover, SNSPDs have contributed to the evolution of interplanetary telecommunications, where quantum com- munication is crucial for securely transmitting information over cosmic distances [129, 269]. Another impactful area is advanced imaging [270]. Their ability to detect events on extremely short timescales, coupled with efficient photon counting, facilitates the capture of high-resolution images in diverse con- texts—from fundamental physics research to medical applications. Studies have showcased their utility in biomedical imaging, including in vivo applications [174, 270, 271].

In Ref. [272–274], SNSPDs were designed with a nanoscale meandering current path etched into a thin layer (less than 10 nm) of superconducting film. These detectors achieve zero resistance when operated at subcritical cryogenic temperatures and with a polarized current below $J_c$. In this configuration, the absorption of thermal energy from a single incident photon or the kinetic impact of a large atom or  ion can momentarily disrupt the superconducting state in the nanowire, creating a localized region of normal resistance. This disruption in current flow generates a radio frequency (RF) "sensing pulse" along the polarization line. The devices are typically mounted in self-aligning fiber-packed holders and cooled within sorption-based cryostats to temperatures ranging from 720 to 780 mK [254].

Superconducting quantum devices, such as SNSPDs and SQUIDs, are driving technological advance- ments that align with SDG 9 (Industry, Innovation, and Infrastructure). These devices significantly enhance the efficiency of existing technologies, such as semiconductor photon detectors, by offering more sensitive and efficient alternatives like SNSPDs. Additionally, both SQUIDs and SNSPDs improve the efficiency and security of applications in information, particularly in areas requiring precise detection and measurement at the quantum level.

Superconducting quantum computers with superconducting qubits make use of the Josephson effect and thus offer an unique distinguishable attribute to other types of qubits realized by conventional conductors. Research on superconducting qubits is carried out by several well-known companies, including IBM and Google. This research makes direct use of the developments carried out in the Josephson technology project at IBM [275], where Josephson logic and memory technology was intended for future high-speed computer systems. Now, for quantum computing, three superconducting qubit architectures exist, the phase, charge (transmon) and flux qubit. Especially the flux qubit makes use of trapping magnetic flux quanta in a superconducting ring. The superconducting qubits represent one of the most mature platforms for quantum information processing [276], and state-of-the-art processors based on superconducting transmon qubits were already scaled up to 65 fully programmable qubits on one chip [277, 278]. So, many recently developed quantum chips use superconducting architecture and require cooling to the milliKelvin range for better stability. The materials of choice for these systems are Nb, Ta and Al as extremely pure materials are required to avoid any scattering at impurities.

## 7.5    Superconductors and the Hydrogen economy

An important issue for a $CO_2$-free world is the use of hydrogen as a fuel. Combusting hydrogen produces only water and no carbon-oxides or nitrogen-oxides are involved in this process. Furthermore, the element hydrogen is abundant in the universe, has the highest gravimetric energy density of 120 MJ/kg, and shows sustainability and environmentally friendly features. As illustrated in Figure 17, the hydrogen economy involves several steps, can be from productioon "green" Hydrogen by elctrolysis from $H_2O$ using sustainable sources of energy like hydropower, wind, solar and nuclear power. For transport and storage of hydrogen, liquefaction of the hydrogen gas is demanded to save volume. $LH_2$ has a three times higher energy density than the compressed hydrogen gas at a pressure of 350 bar.

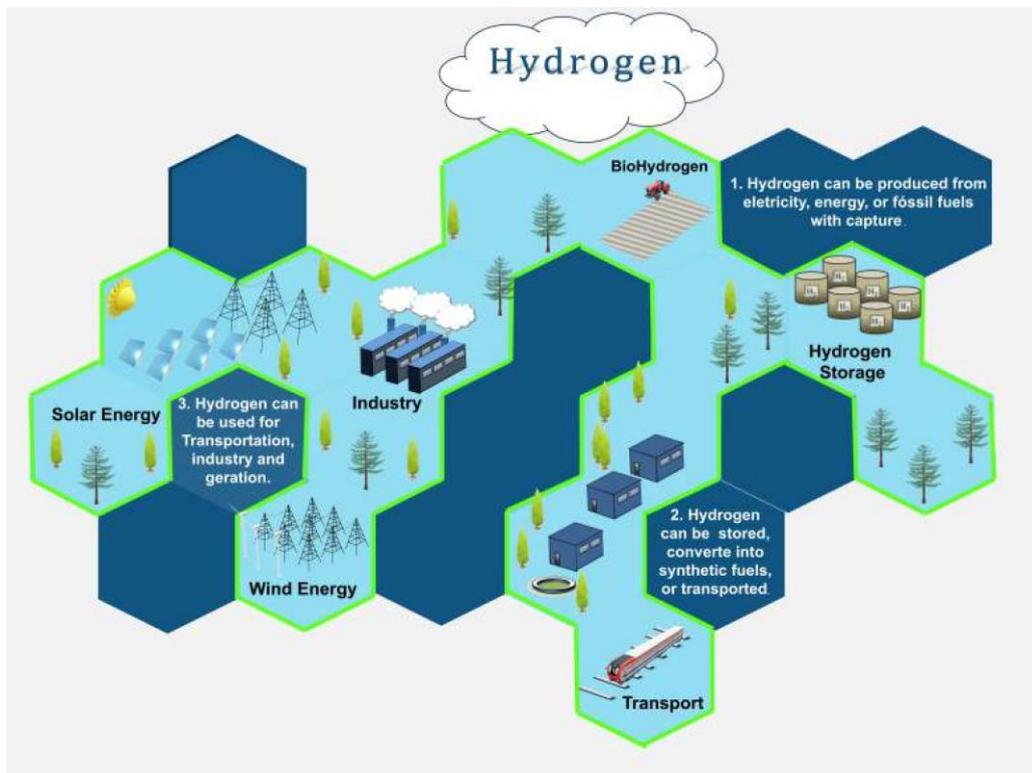

Figure 17: The hydrogen economy

The boiling temperature of $LH_2$ is 20.3 K. To enable an effective use of $LH_2$, superconductivity comes into play: The presence of $LH_2$ allows an effective synergy with the cooling of superconducting wires, so the dreams of a superconducting pipeline [279] may come true. If one can insert a very compact medium-voltage direct-current superconducting cable into an $LH_2$ pipeline, this would enable an unique solution to simultaneously distribute the two energy vectors electricity and hydrogen. On a smaller scale, the use of $LH_2$ for fuel cells may enable the use of a superconducting generator or propulsion system coupled by an heat exchanger. Thus, superconductivity has the potential to reduce the costs of $LH_2$ by comsuming less energy.

The materials of choice for the $LH_2$ are, as already mentioned in Section 7, the metallic superconductor $MgB_2$ and the IBS 122 materials, both of which having $T_c$ being double the temperature of $LH_2$ and being cheap as there are no rare earths involved in their production. Thus, the hydrogen economy, which is part of many recent efforts to develop a de-carbonized world (e.g., the European Green Deal), can develop its full potential with the combination of superconducting cables and devices.

# 8 Authors' perspectives

This review employs bibliometric methodology to present the first comprehensive bibliometric analysis of superconductors in relation to the UN Sustainable Development Goals (SDGs). While the review encompasses a global bibliometric analysis of superconductor applications from 1986 to 2023, our primary focus is an in-depth examination of superconductors and their applications toward the implementation of the SDGs between 2015 and 2023. Specifically, we provide a detailed bibliometric analysis of leading journals, countries, institutions, and the most relevant articles. Additionally, we aim to establish complex relationships between superconductivity applications and the SDGs using meta-analysis strategies such as global collaboration networks, keyword analysis, thematic focus, and identification of hot topics.

Since the discovery of high-temperature superconducting materials, the volume of publications related to their applications has grown steadily, with a notable increase in the number of articles and citation rates since the adoption of the Sustainable Development Goals (SDGs) in 2015. For articles concerning superconductivity and its applications, we observe a distribution of publications across various SDGs, with a primary focus on SDG 07 (Affordable and Clean Energy), SDG 03 (Good Health and Well- being), and SDG 09 (Industry, Innovation, and Infrastructure). As the 2030 Agenda has progressed, the impact of articles on superconductivity and its applications has become more balanced, with contributions affecting almost all SDGs to varying degrees. This shift reflects a growing trend toward interdisciplinary and transversal research within the framework of the UN 2030 Agenda. However, it is evident that the superconductivity community is not consistently using keywords or text that clearly highlight the potential importance of their work for the SDGs.

Our analysis also revealed that more developed countries, such as China, the USA, and Japan, occupy the top three positions among the most productive countries in the field. Additionally, both the G7 and BRICS countries are equally influential in superconductivity and its applications, with BRICS' influence being primarily driven by China. These countries concentrate the most prominent institutions during the implementation period of the SDGs. Conversely, G20 member countries and developing nations require more investment in this field. G7 countries generally have more established and globalized collaboration networks, which could serve as a focal point for BRICS governments (excluding China). Developing countries should emphasize research aligned with the SDGs to contribute effectively to the global 2030 Agenda.

There is a noticeable gap between general and specialized journals in terms of impact based on cita- tions. Superconductivity scholars tend to publish in more traditional journals such as IEEE Transactions on Applied Superconductivity, Journal of Superconductivity, Journal of Superconductivity and Novel Magnetism, and Physical Review B. In contrast, the most influential articles on superconductivity and the SDGs are published in broad-scope and high-impact journals, such as Nature Materials, Nature Physics, and Advanced Materials. To effectively contribute to the achievement of the UN SDGs, super- conductivity researchers should consider directing their work toward journals with a broader scope and aligned goals. Embracing transdisciplinarity represents the most efficient path to fulfilling the SDGs.

The analysis of thematic focus and hot topics revealed a prominent role for non-conventional su- perconductors, particularly YBCO [140], iron-based superconductors [80], and $MgB_2$ [164]. However, new materials such as nickelates [118] are increasingly capturing the community's attention. From the SDGs' perspective, ceramic superconductors have been studied for applications in MRI technology, en- abling smaller devices with minimal maintenance and simplified cooling requirements [179–181]. In mass transport, superconducting Maglevs remain a long-term challenge, but we highlight the potential sub- stitution of ship engines with electric superconducting ones. In nanotechnology, notable studies focus on using ceramic superconductors as superconducting nanowire single-photon detectors (SNSPDs) [205], with potential new morphologies achievable through low-cost techniques like electrospinning [280, 281] and solution blow spinning [282, 283].

The distribution of influential articles on the applications of superconductors linked to the SDGs is driven by the transition from fossil fuels to more sustainable energy sources, such as the hydrogen econ- omy, as part of the broader shift toward a sustainable economy. The future of superconducting materials for sophisticated and expensive technological applications may depend on their ability to integrate into this emerging hydrogen economy. In this context, the use of hydrogen (e.g., liquid $H_2$ as refrigerant) could enhance the viability of developments such as superconducting electric transmission lines in con- junction with liquid hydrogen transmission, electric trains (and MagLevs) powered by green $H_2$, and novel sustainable data centers for quantum computers using advanced superconducting materials with liquid $H_2$ as a refrigerant. All these developments contribute to a broader superconducting and green $H_2$ commodity market.

# 9 Acknowledgments

We acknowledge the Brazilian agencies São Paulo Research Foundation, FAPESP, grants 2016/12390- 6 and 2021/08781-8, Coordenação de Aperfeiçoamento de Pessoal de N´ivel Superior- Brazil (CAPES) - Finance Code 001, UNESP-CAPES/PrInt program, and National Council of Scientic and Technological Development (CNPq, grant 310428/2021-1), R.I. thanks the UNESP for the financial support through

CAPES/PRINT - Edital nº 41/2017, UNESP nº 5089, UNESP nº 5089 and 5530.

# 10 Authors' Credits

- **EASD:** Methodology, Validation, Formal analysis, Data curation, Investigation, Writing - Original Draft Preparation, Writing - Review & Editing.

- **AEAP:** Validation, Formal analysis, Data curation, Investigation, Writing - Original Draft Prepa- ration, Writing - Review & Editing.

- **RI:** Conceptualization, Methodology, Validation, Formal analysis, Data curation, Writing - Original Draft Preparation, Writing - Review & Editing.

- **MRK:** Conceptualization, Investigation, Supervision, Writing - Review & Editing.

- **AKV:** Investigation, Writing - Review & Editing.

- **DMK:** Visualization, Writing - Review & Editing.

- **RZ:** Conceptualization, Writing - Review & Editing, Visualization, Supervision, Project adminis- tration, Funding acquisition.